\documentclass[fleqn,usenatbib]{mnras}

\usepackage{newtxtext,newtxmath}

\usepackage[T1]{fontenc}
\usepackage{ae,aecompl}

\usepackage{graphicx}	
\usepackage{amsmath}	
\usepackage{amssymb}	
\usepackage{hyperref}
\newcommand{\survey}{K2:BS}
\newcommand{\name}{KSN:BS-C11a}

\newcommand{\kepler}{{\it K2/Kepler}}

\title[New CV in {\it K2/Kepler}]{Discovery of a New WZ~Sagittae Type Cataclysmic Variable in the \textit{Kepler/K2} Data}

\author[R. Ridden-Harper et al.]{R. Ridden-Harper,$^{1,2}$\thanks{E-mail: ryan.ridden-harper@anu.edu.au}
B. E. Tucker,$^{1,2,3}$
P. Garnavich,$^{4}$
A. Rest,$^{5,6}$
S. Margheim,$^{7}$\newauthor 
E. J. Shaya,$^{8}$
C. Littlefield,$^{4}$
G. Barensten,$^{9}$
C. Hedges,$^{9}$
and M. Gully-Santiago$^{9}$
\\
$^{1}$Research School of Astronomy \& Astrophysics, Mount Stromlo Observatory, The Australian National University,\\ Cotter Road, Weston Creek, ACT 2611, Australia\\
$^{2}$The ARC Centre for All-Sky Astronomy in 3-Dimensions (ASTRO3D).\\
$^{3}$The National Centre for the Public Awareness of Science, The Australian National University, Canberra, ACT 2601, Australia.\\
$^{4}$Physics Department, 225 Nieuwland Hall, University of Notre Dame, Notre Dame, IN 46556, USA.\\
$^{5}$Space Telescope Science Institute, 3700 San Martin Drive, Baltimore, MD 21218, USA.\\
$^{6}$The Johns Hopkins University, Baltimore, MD 21218, USA.\\
$^{7}$Gemini Observatory, c/o AURA, Casilla 603, La Serena CL.\\
$^{8}$Astronomy Department, University of Maryland, College Park, MD 20742-2421, USA.\\
$^{9}$NASA Ames Research Center
.
}

\date{Accepted XXX. Received YYY; in original form ZZZ}

\pubyear{2019}

\begin{document}
\label{firstpage}
\pagerange{\pageref{firstpage}--\pageref{lastpage}}
\maketitle

\begin{abstract}
We identify a new, bright transient in the \kepler\ Campaign 11 field. Its light curve rises over seven magnitudes in a day and then declines three magnitudes over a month before quickly fading another two magnitudes. The transient was still detectable at the end of the campaign. The light curve is consistent with a WZ~Sge type dwarf nova outburst. Early superhumps with a period of 82 minutes are seen in the first 10 days and suggest that this is the orbital period of the binary which is typical for the WZ~Sge class. Strong superhump oscillations develop ten days after peak brightness with  periods ranging between 83 and 84 minutes. At 25 days after the peak brightness a bump in the light curve appears to signal a subtle rebrightening phase implying that this was an unusual type-A outburst. This is the only WZ~Sge type system observed by \kepler\ during an outburst. The early rise of this outburst is well-fit with a broken power law. In first 10 hours the system brightened linearly and then transitioned to a steep rise with a power law index of 4.8. Looking at archival \kepler\ data and new \textit{TESS} observations, a linear rise in the first several hours at the initiation of a superoutburst appears to be common in SU~UMa stars. 
\end{abstract}




\section{Introduction} \label{sec:intro}

Cataclysmic variable stars (CVs) consist of a white dwarf star accreting material from a companion star. These accreting binary systems show a rich array of physical processes that are displayed over many different timescales \citep[for a review, see][]{Warner1986}. Mass transfer from the companion star is often facilitated through an accretion disk around the white dwarf star. Thermal instability in the disk drives outbursts known as dwarf novae \citep[DN;][]{Osaki1974, Hoshi1979}. The character of these DN outbursts depends on the binary orbital period. 

Accretion disks in systems with orbital periods less than two hours can experience periodic ``superoutbursts'' that are more luminous and last longer than normal outbursts. This class of DN are named for their prototype, SU~Ursa~Majoris (SU~UMa). During a superoutburst, the disk can be excited by a 3:1 orbital resonance with the binary orbit that generates strong optical oscillations known as ``superhumps'' (SHs) \citep{Whitehurst1988, Osaki1989, Lubow1991, Lubow1991a}. The frequency and amplitude of the SHs evolve over the outburst in specific ways that are good diagnostics of the physics in the disk and binary system.

SU~UMa DN are thought to be evolving to shorter orbital periods through energy loss by gravitational radiation. As the orbital periods shorten, the secondary mass decreases and the mass ratio of the systems become extreme. At periods around 80 minutes, the separation between the components begins to increase in what is known as the period bounce \citep[][and references therin]{kolb1999, Knigge2011}. CVs near the bounce period have mass ratios less than 0.1 and these are known as WZ~Sge type DN. \citet{Kato2015} presents an extensive review of these binaries. WZ~Sge stars have a number of distinctive properties, such as very large amplitude outbursts, brightening by 7 or 8 magnitudes. The time between outbursts can be many years to decades, meaning that catching a particular system in a bright state is rare. Unlike other SU~UMa DN, WZ~Sge DN feature double-peak variations in their light curves known as ``early superhumps''. These early superhumps are a result of the high mass ratio of WZ~Sge systems, that allows the accretion disk to reach the 2:1 orbital resonance \citep{Osaki2002}. Early superhumps are a valuable diagnostic tool, featuring periods that correspond closely with the orbital period \citep{kato2002,ishioka2002}.

Another feature common to WZ~Sge DN is a late time rebrightening event at the end of a superoutburst. As described in \citet{Kato2015}, these events have been  classified into 5 categories: type--A/B outbursts, long duration or multiple rebrightenings; type--C outbursts, single rebrightening; and type--D outbursts, no rebrightening, all identified by \citep{Imada2006}; and type--E, double superoutburst, identified by \citet{kato2014}. Using the rebrightening types \citet{Kato2015} constructs an evolutionary sequence that can be used to identify the age of the system and the configuration, with the evolutionary sequence type C~$\rightarrow$~D~$\rightarrow$~A~$\rightarrow$~B~$\rightarrow$~E. Type--A outbursts were found to be close to the period minimum, making them good period bouncing candidates. 

The \textit{Kepler} space telescope has proven to be a valuable instrument in detecting and understanding DN. With the 1~minute short cadence, and 30~minute long cadence data, \textit{Kepler} has obtained exquisite light curves of known and previously unknown DN.  \citet{kato2013b} presents the analysis of three SU~UMa DN, V585~Lyr and V516~Lyr and the serendipidously discovered KIC~4378554 presented in \citet{Barclay2012}. With the high photometric precision of \textit{Kepler} data and long baseline afforded by the \textit{Kepler} prime mission, \citet{kato2013b} presented evidence that supported the thermal--tidal instability theory as the origin of superoutbursts.

Like {\it Kepler}, {\it K2} data offers the unparalleled high cadence, 30~minute observations of thousands of targets \citep{Borucki2010,Koch2010,Howell2014}. K2 differs from \textit{Kepler} in that each science target is allocated a larger detector area to compensate for telescope drift, as a result fewer science targets were observed. Although drift reduces the photometric precision of {\it K2}, the increase in contiguous background pixels improves the chances that transients can be detected.

Here, we present observations of a new WZ~Sge type CV discovered as part of a systematic search for new transients in the \kepler\ campaigns, known as the K2: Background Survey (K2:BS). Our new CV is the first WZ~Sge type system to be observed with the high cadence and continuous monitoring of \textit{Kepler}. In Sec.~\ref{sec:data} we describe the detection method used to discover \name. Furthermore, we outline the \kepler\ light curve processing and spectroscopy of the system. In Sec.~\ref{sec:analysis} we analyse the \kepler\ light curve and the unique high cadence coverage of the rise to peak brightness. Finally in Sec.~\ref{sec:discussion} we compare the light curve of \name\ with other WZ~Sge stars to derive fundamental parameters of the binary system.

\section{Data} \label{sec:data}

\subsection{Search for Transients in \kepler}
Through the \kepler\ Extra-Galactic Survey (KEGS), \kepler\ observed thousands of galaxies and provided a wealth of information on transients that were detected in scheduled galaxy targets \citep[e.g.][]{Garnavich2016, Rest2018, Dimitriadis2018}. Although many transients were detected in the directed search, there are more transients hidden in the \kepler\ data. Each science target in \kepler\ has numerous background pixels that are observed at high cadence. During a campaign these background pixels may serendipitously collected transient signals, which have gone previously undetected.

\citet{Ridden-Harper2020} presents the K2: Background Survey (\survey ), which conducts a systematic search for transients in \kepler\ background pixels. \survey\ independently analyses each pixel to detect abnormal behaviour. This is done by searching for pixels that rise above a brightness threshold set from the median brightness and standard deviation through a campaign. Telescope motion presents a challenge in candidate detection as science targets may drift into background pixel, triggering false events. False triggers are screened by vetting of events that last $<1$~day, chosen for candidates with the 6 hourly telescope resets. Coincident pixels that pass the vetting procedure are grouped into an event mask. All candidate events are checked against the NASA/IPAC Extragalactic Database (NED)\footnote{The NASA/IPAC Extragalactic Database (NED) is operated by the Jet Propulsion Laboratory, California Institute of Technology, under contract with the National Aeronautics and Space Administration.} and the SIMBAD database \citep{Wenger2000} to identify potential hosts. \survey\ produces event figures, videos and relevant detector information, which are used for manual vetting.

\subsection{Discovery of \name }

A bright transient, \name, was detected in the \kepler\ archived target pixel file (TPF) for EPIC 203830112 Campaign 11 (C11). The science target EPIC~203830112 was proposed for observation by several programs in C02 and C11 shown in Table~\ref{tab:proposals}. The science target is a F-type dwarf that was selected for magnitude limited surveys of FGK dwarf stars with the objective of understanding the occurrence rate and properties of near-Earth sized planets. The location of \name\ was observed in both C02 and C11, however, it was in quiescence in C02.

\name\ was discovered in \kepler\ C11 galactic pointing campaign, during a preliminary search of \survey. This event went undetected as transient surveys \citep[e.g.][]{Rest2018} avoided the crowded galactic fields. Furthermore, \name\ was unlikely to be detected from ground based surveys as \textit{Kepler} was ``backward facing'' during C11, meaning that the target field appeared in the evening twilight as viewed from the Earth. The discovery of \name\ shows the capability of \survey\ to rapidly analyse \kepler\ data for transients, even in galactic pointing fields.

\begin{figure*}
	\includegraphics[width=\textwidth]{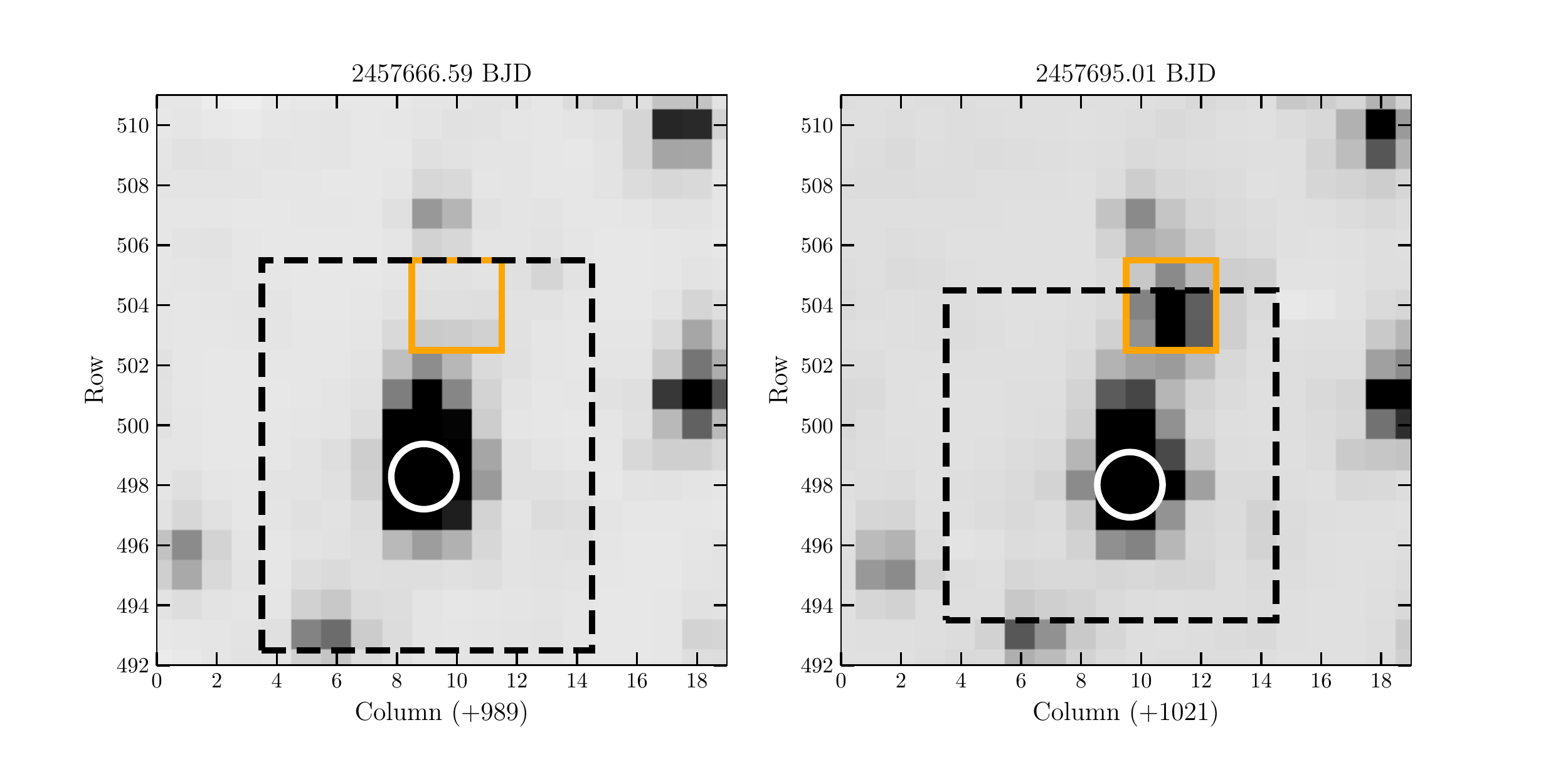}
	\caption{\kepler\ Full Frame Images (FFIs) for C11, the white circle indicates the science target for EPIC 203830112, the orange solid box indicates the chosen event aperture, and the black dashed boxes indicate the K2 target pixel file boundaries for the C11 sub-campaigns (C111 and C112). The C111 image is shown on the left and the C112 on the right, which contains \name\ $\sim$3 days post max.} \label{fig:FFIs}
\end{figure*}

\subsection{Full Frame Images}
In each \kepler\ campaign, at least one full frame image (FFI) is taken. Many campaigns, such as C11, have two FFIs: one at the start, and one midway through the campaign. In rare cases when a transient occurs during an FFI, it can be shown in relation to all other objects in the region. Such occurrences are useful to establish the reality of the transient and determine its total brightness.

\begin{figure}
	\includegraphics[width=\columnwidth]{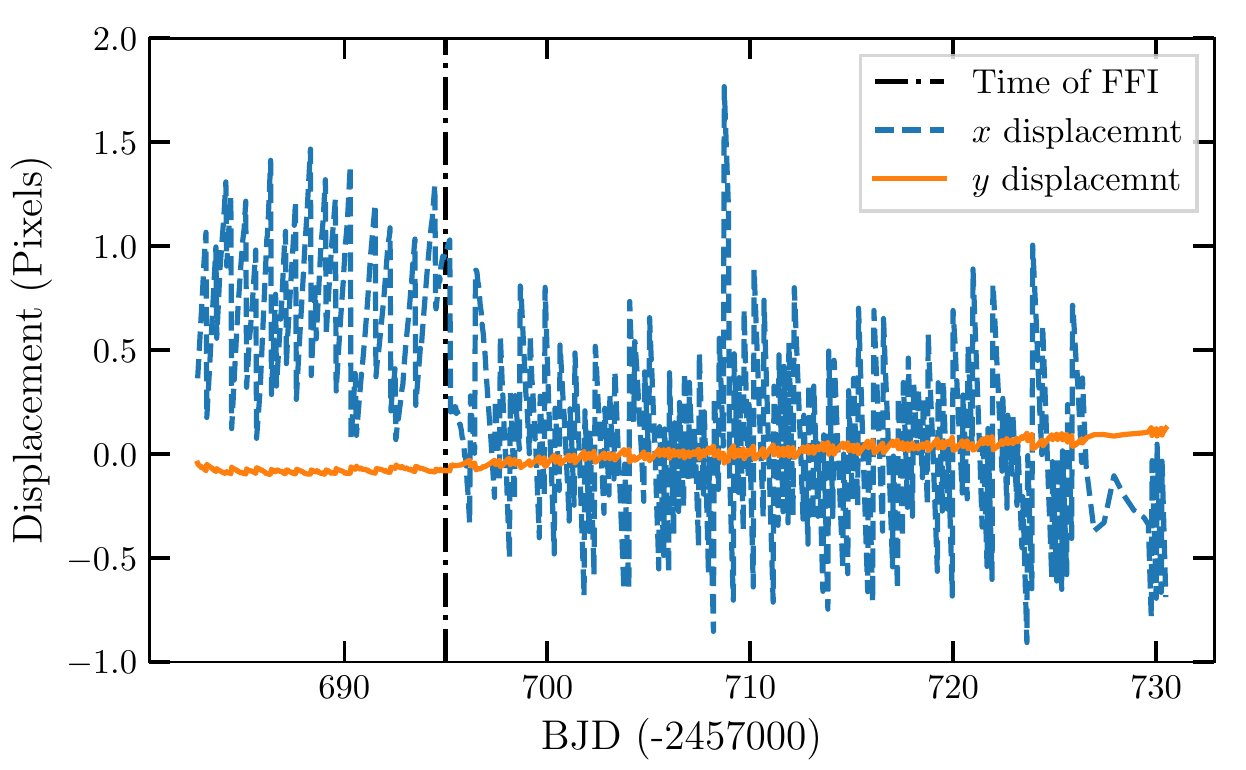}
	\caption{Kepler's displacement during the C112 sub-campaign. Although there is large $x$-displacement (across pixel columns), there is negligible $y$-displacement (across pixel rows), only deviating by 0.1~pixels during the event. It seems reasonable that the TPF consistently receives $\sim$96\% of the event flux.}\label{fig:displacement}
\end{figure}

The two FFIs for C11 were taken at 2457666.59~BJD and 2457695.01~BJD, where the second FFI occurs $\sim$3 days after \name\ reached peak brightness. As shown in Fig.~\ref{fig:FFIs} the event is well separated from the science target, by $\sim3$~pixels, or $\sim12''$. The presence of \name\ in the FFI is strong confirmation that \name\ is not an artefact of the data.

Because the transient is at the edge of the TPF, the wings of the point-spread-function (PSF) are cutoff in the individual images. We can use the FFIs to estimate the fraction of event flux that the TPF missed. Aperture photometry was preformed with the event apertures shown in Fig.~\ref{fig:FFIs} with the C111 sub-campaign aperture for background subtraction. The FFI counts were $C_{FFI} = 20586~e/s$, while 
the event light curve recorded lower counts of $C_{lc}=17556~e/s$. Thus, the TPF received $\sim$96\% of the total event flux. As noted below, we find that there is little spacecraft motion that would move \name\ off the TPF, so it likely that a high level of flux was captured for the entire event.


\begin{figure}
	\includegraphics[width=\columnwidth]{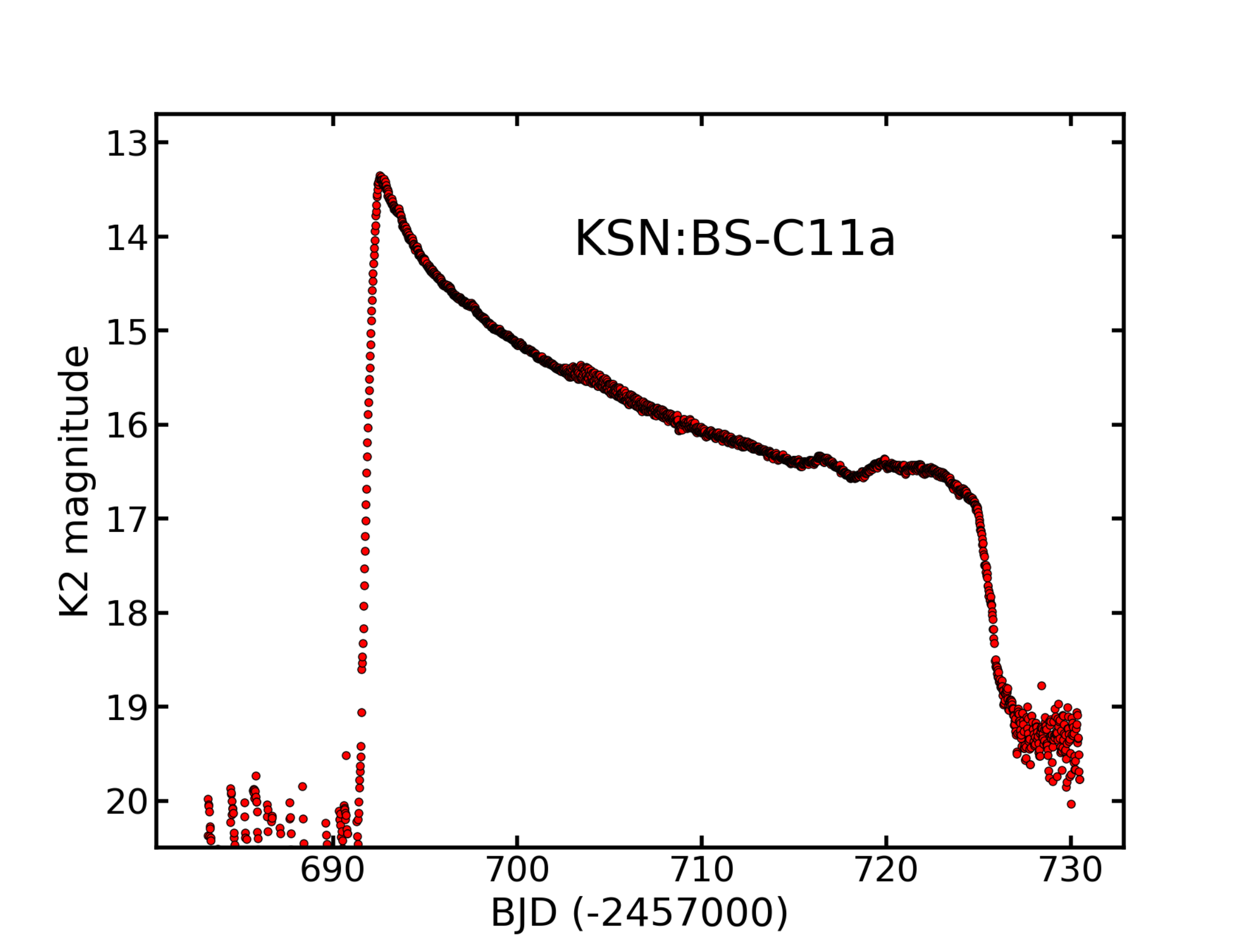}
	\caption{The \kepler\ light curve of the transient \name\ shown with the 30-minute cadence. The time axis is shown in barycentric Julian days and the flux has been converted to \textit{Kepler} magnitudes.
		\label{fig:lightcurve}}
\end{figure}

\subsection{\textit{K2/Kepler} Light Curve}

The light curve for \name\ was constructed through simple aperture photometry. Due to telescope motion across the TPF, as seen in Fig.~\ref{fig:displacement}, the custom aperture was chosen to extend across many columns, to encapsulate all flux. Although a larger mask introduced larger background noise, this was inconsequential due to the high event count rate, and a quiescent period at the beginning of sub-campaign C112, which is ideal for background subtraction.

A less successful method of data reduction was attempted using Point Response Function (PRF) model fitting photometry. This data reduction method is available though the Lightkurve python package\newline \citep{lightkurve}, in which a PRF is fit to each frame to determine telescope properties and recover the flux, free from telescope effects. Although this method proved successful in accounting for telescope motion, it was unable to recover the total counts of the event, thus the aperture photometry method was favoured. 

The light curve from the \kepler\ data is shown in Fig.~\ref{fig:lightcurve}.
Due to the high count rate of this event, and the rapid cadence of \kepler\ data, the light curve of the event is extremely well defined. The event had a rapid rise of $> 7$~mag over $\sim$1~day, followed by a slow decay lasting about a month. Large amplitude oscillations began about 10~days after maximum light. Overall, the light curve appears to be that of a short period CV in superoutburst. In particular, its characteristics are similar to outbursts of GW Librae \citep{vican2011}, a WZ~Sge type CV. 



\subsection{Gemini Spectra}

Following the discovery of \name, Pan-STARRS1 (PS1) images of the field were searched to identify the system in quiescence. Given the 4'' pixel-scale of the \kepler\ data, however, there were several possible progenitors within the uncertainty ellipse. Spectra of the first candidate was obtained on 2018 July 20 (UT), the remaining two candidates were observed on 2018 July 25 (UT) (MJD58324.3) with Gemini Multi-Object Spectrograph North (GMOS-N), using the R400 grating with a wavelength range of 3960--8687~\AA and a 1'' longslit. The data were processed using standard routines within the Gemini IRAF package. The faintest candidate revealed a blue continuum with a broad, double-peaked H$_\alpha$ emission feature which confirmed this source as the dwarf nova.

\begin{figure}
	\includegraphics[width=\columnwidth]{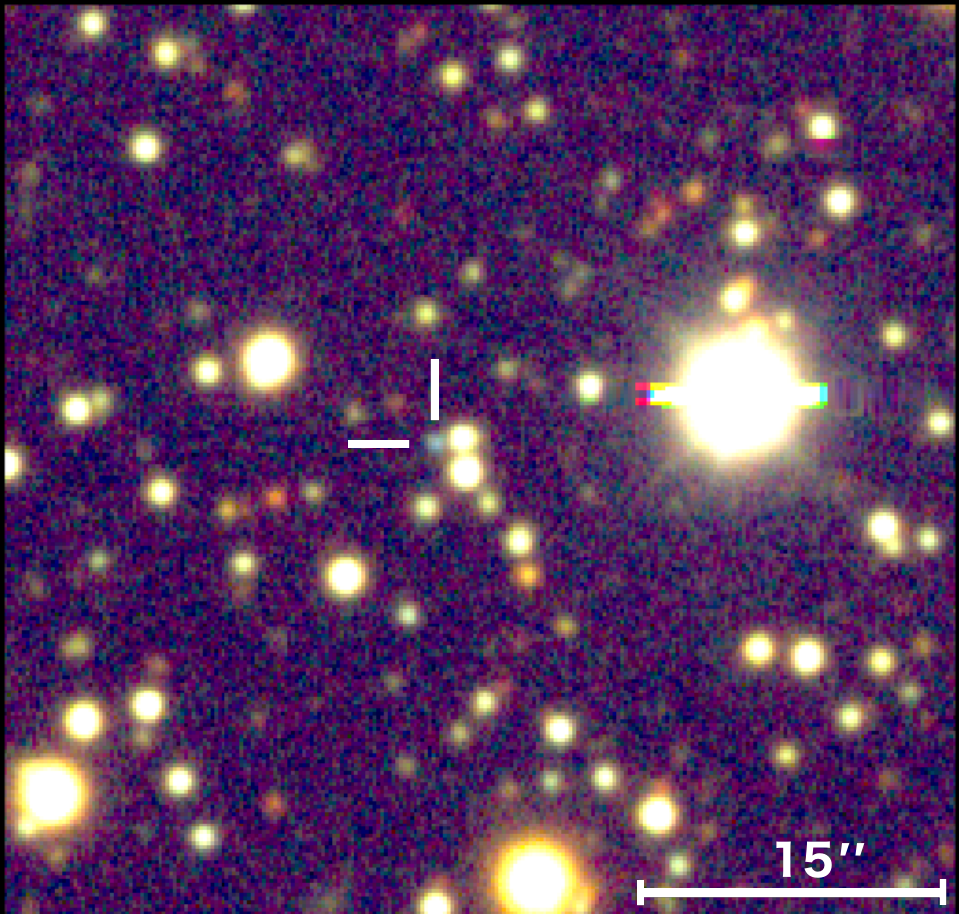}
	\caption{A pseudo-color image of \name\ constructed from $g$, $r$, and $i$ filter images obtained with DECam on the CTIO 4.0~m telescope. The images were taken on MJD 58362.9 when \name\ was in quiescence. The variable is the the blue object marked by dashes. North is toward the top and east is to the left in the image.  \label{fig:color}}
\end{figure}

Using positions for nearby stars from the USNO-B1.0 catalog, we measure the astrometric position of the progenitor to be $\rm RA=16:53:50.67$ $\rm DEC=-24:46:26.50$ (J2000) with an uncertainty of 0.24 arcsec. The location of the star in the quiescent state is shown in Fig.~\ref{fig:color}.

\subsection{DECam images}

Following spectroscopic confirmation with Gemini, deep images were taken with CTIO 4-meter telescope and the DECam instrument on 2018 September 1 (UT) (MJD58363.0). \name\ was imaged in four SDSS filters, with exposure times of 120~s for $g$ and $r$, and 90~s for $z$ and $i$. The images were reduced following standard procedure with the DECam pipeline. 

Photometry in the crowded field was performed by point-spread-function (PSF) fitting with DAOPHOT implemented in IRAF\footnote{IRAF is distributed by the National Optical Astronomy Observatory, which is operated by the Association of Universities for Research in Astronomy (AURA) under a cooperative agreement with the National Science Foundation.}. An average PSF was constructed from isolated stars in the field and this function was scaled to match the brightness of the two stars just west of the variable. After subtracting these stars, aperture photometry was performed on \name\ and on nearby stars calibrated in the PS1 Survey DR1 catalogue \citep{Chambers2016}. We find the quiescent brightness of \name\ is $r=21.46\pm 0.08$ and
$g=21.52\pm 0.07$. From these estimates we approximate the magnitude in the \textit{Kepler} system as $K_p = 21.5$. Given that the \textit{Kepler} brightness at the peak of the outburst was $K_p \sim 13.4$, the full amplitude of the outburst was 8.1~mag.

\section{Analysis} \label{sec:analysis}
\subsection{Power Spectrum} \label{sec:powerspec}
\begin{figure*}
	\includegraphics[width=\textwidth]{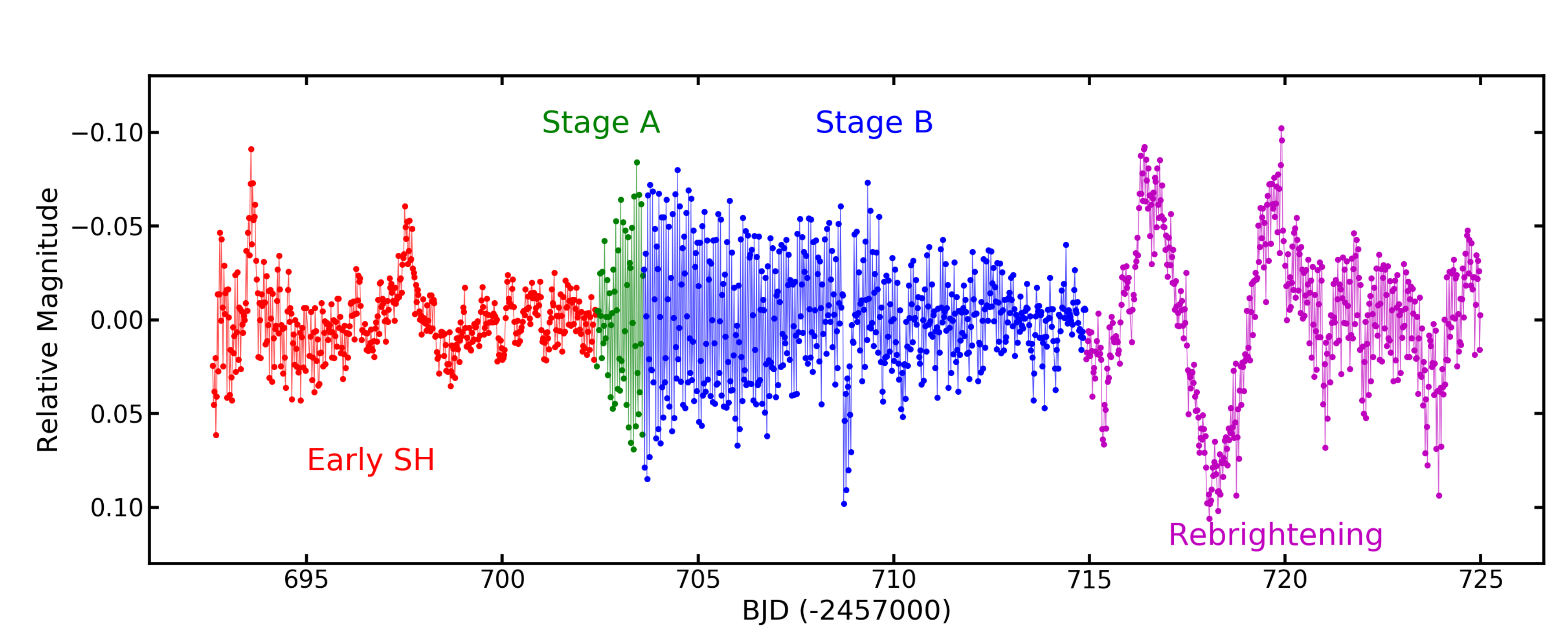}
	\caption{The outburst light curve starting at maximum light with the fading trend subtracted. The initial section (red) shows the Early SHs. The development of the standard SHs, or Stage A is shown in green. The Stage B SHs decline in amplitude and frequency (blue). Finally, a rebrightening stage is shown in magenta.  \label{fig:residuals}}
\end{figure*}

During a superoutburst, oscillations occur in the light curve caused by orbital resonances so that the power spectrum of an outburst can probe the physics of the disk. The high cadence observations of \kepler\ are capable of identifying the short scale superhump oscillations and measure their time evolution \citep{Barclay2012,Kato2013,Kato2013a,Brown2015b}. To study the oscillations during the outburst we have removed the slow decline from maximum by subtracting a median-filtered light curve with using a 36~hour wide kernel. The resulting residual light curve is shown in Fig.~\ref{fig:residuals}.

\begin{figure}  
	\includegraphics[width=\columnwidth]{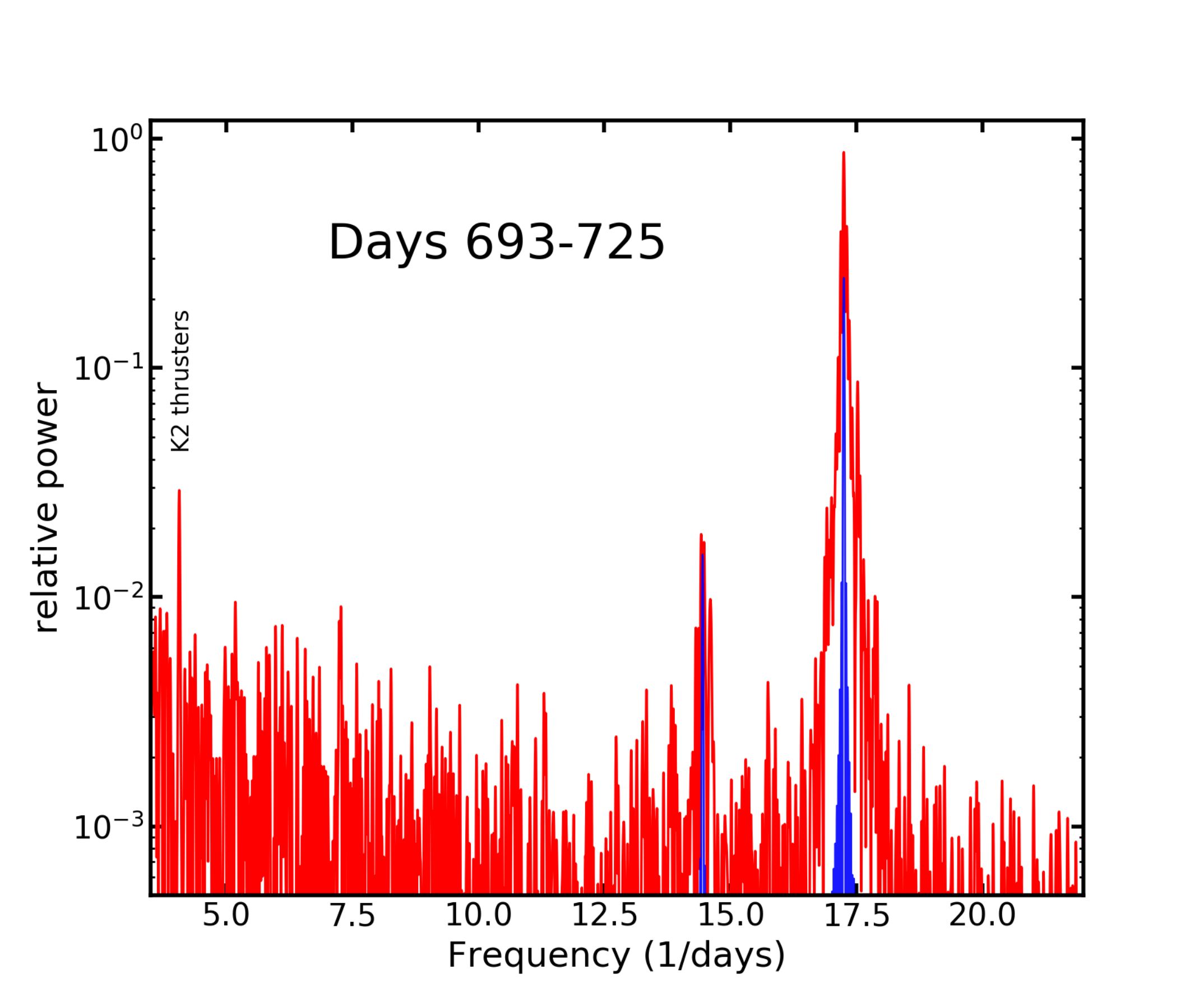}
	\caption{The power spectrum of the light curve over the entire outburst (red line). For comparison, the power spectrum of a periodic function with a frequency of 17.25~day$^{-1}$ is plotted in blue to demonstrate that the power near 14.4~day$^{-1}$ results from a harmonic beyond the Nyquist frequency. The peak near 4~day$^{-1}$ comes from the regular \kepler\ thruster firings. \label{fig:pow}}
\end{figure}

The evolution of a WZ~Sge outburst begins with ``early'' superhumps that tend to have a low amplitude and a period close to the orbital period of the system. An increasing amplitude signals the Stage~A superhumps that rapidly transition to the Stage~B. The outburst finally shows variations on the time scale of a few days that we associate with the rebrightening phase \citep{Kato2015}.

The power spectrum of \name\ over the outburst is seen in Fig.~\ref{fig:pow}, and reveals two broad peaks at frequencies around 17.2~day$^{-1}$ and 14.4~day$^{-1}$. A narrow peak at 4~day$^{-1}$ is due to the regular 6~hour spacecraft drift plus thruster correction cycle.

The power at 14.4~day$^{-1}$ is likely the second harmonic of the primary peak sampled beyond the \textit{Kepler} long cadence Nyquist frequency \citep{Barclay2012}. This results in a folded alias of that harmonic and this is not a true periodicity in the light curve. To test this, we took a power spectrum of a periodic function with a frequency of 17.25~day$^{-1}$. The power spectrum of the periodic function had significant harmonics at 34.5~day$^{-1}$ when sampled at a fast cadence. When re-sampled at the \kepler\ cadence a folded alias appeared at 14.4~day$^{-1}$, and this alias plus the true power peak are shown in blue in Fig.~\ref{fig:pow}.

\begin{figure}
	\includegraphics[width=\columnwidth]{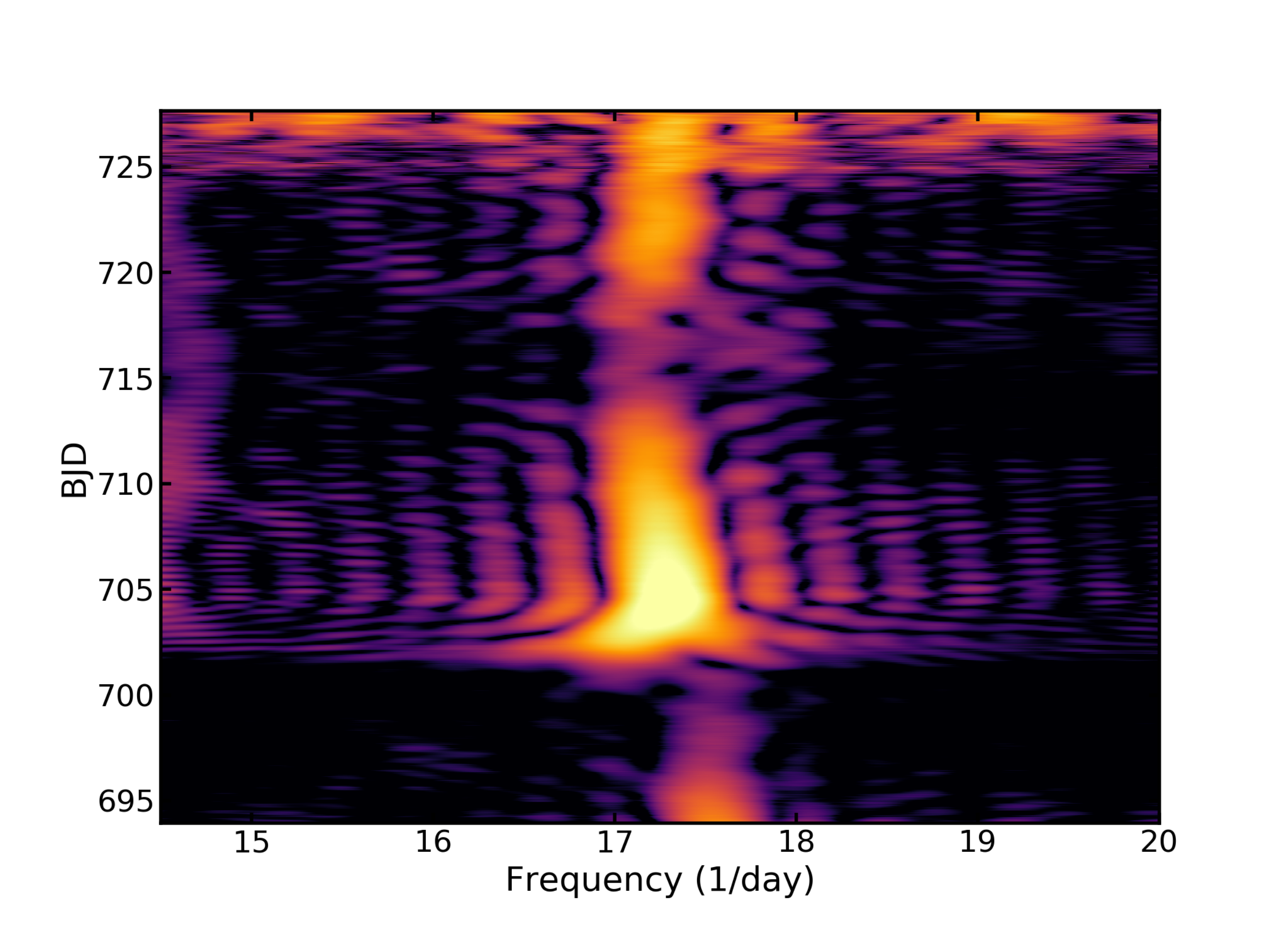}
	\caption{The time-resolved power spectrum of the light curve starting at the peak of the outburst and plotted against BJD ($-2457000$). Each power spectrum is calculated from a moving 2.7~days of data. The strongest peak shows an Airy pattern (sidelobes) due to the sharp edges of the sample box. Early superhumps are most prominent in the first few days, but extend into the Stage~A superhumps phase. Between day 703 and 715 the Stage~B superhumps decrease in frequency.  \label{fig:power}}
\end{figure}

The variation in the superhump frequencies can be seen time-resolved power spectrum shown in Fig.~\ref{fig:power}. Here the early superhumps are seen to have a nearly constant frequency before the transition to Stage~A around day 702. The Stage~B superhumps drift to lower frequencies before fading in amplitude around day 715. 

\subsection{Early Rise}

The photometric cadence of \kepler\ permits the study of the earliest phases in DN outbursts. \kepler\ has observed only a handful of DN superoutbursts and no outbursts of a WZ~Sge star until now.

The fixed aperture \kepler\ photometry of \name\ shows sawtooth pattern with an amplitude of $\pm 100$~e/s. The sawtooth before the outburst is due to the drifting motion of $Kepler$ spacecraft bringing the wings of nearby stars into the aperture. For this target the \kepler\ motion results in image movement mainly along the lines of the CCD (see Fig.~\ref{fig:displacement}). So there is an excellent correlation between the flux variations in the aperture and the $x$-direction image displacement. We fit the correlation with a third order polynomial and subtract the fit from the light curve. After removing a slow trend in the flux, the corrected light curve has a standard deviation about zero of $\sigma = 10.5$~$e/s$ over the seven days before the outburst. 

\begin{figure}
	\includegraphics[width=\columnwidth]{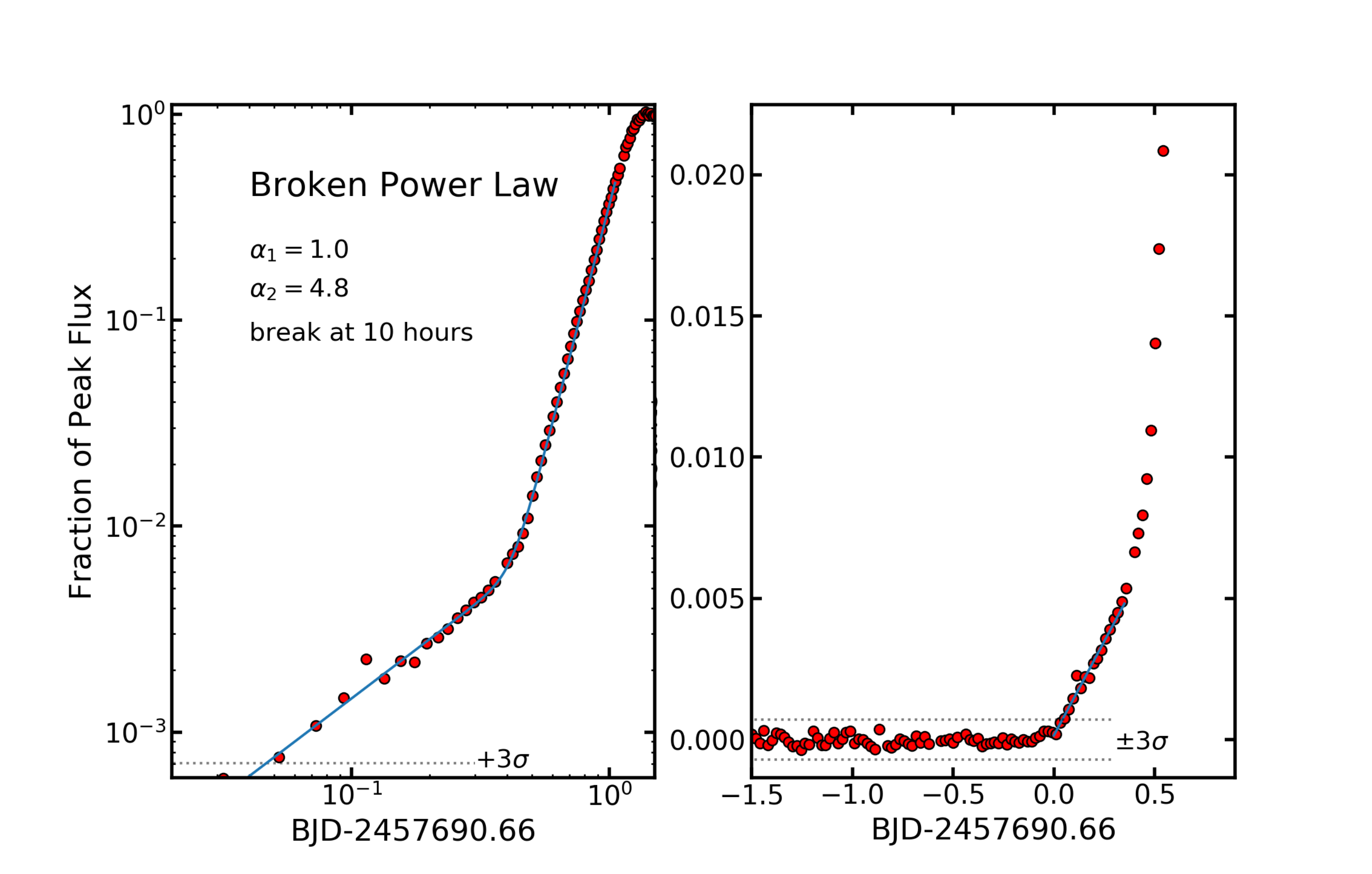}
	\caption{The rising part of the light curve plotted as a fraction of the peak flux. {\bf Right:} The outburst begins slowly before a rapid rise 10 hours after first light. Extrapolating the nearly-linear early light gives a time for the onset of the outburst as BJD2457690.66. The dotted lines indicate the flux corresponding to $\pm 3$ standard deviations of the pre-outburst background counts. {\bf Left:} The first day of the outburst the transient reaches 50\%\ of its peak luminosity and the light curve is well fit by a broken power law. Initially the power law index is 1.0, but after ten hours the index steepens to an index of 4.8. At the beginning of the second day, the rise rate slows from the steep power law and the outburst finally peaks 33~hours after the onset of the burst.   \label{fig:rise}}
\end{figure}

First light of the outburst is detected at BJD2457690.71 when the flux within the aperture rises above three standard deviations of the background. As shown in Fig.~\ref{fig:rise}, the brightness rises slowly over the initial 10~hours after detection. When the system has reached about 0.5\%\ of peak luminosity the rate of brightening dramatically increases and it takes just 23~hours to brighten by the remaining 99.5\%.


A two-component or ``broken'' power law fit provides an excellent description of the rising light curve up to 50\%\ of the peak flux. As expected, the initial linear phase corresponds to a power law with index $\alpha_1 = 0.95\pm 0.09$ followed by a quick transition to a steeper index of $\alpha_2 = 4.82\pm 0.07$ after 10 hours. The rate of brightening begins to slow and deviate from the steepest power law rise after 25~hours.

This initial linear rise has never been observed in a WZ~Sge type star before. The transition to the steep rise occurs around 5.7 magnitudes fainter than the peak, or approximately a \textit{Kepler} magnitude of 19.1~mag. For a quiescent $r= 21.5$ magnitude, the star rose by a factor of 10 in brightness during the slow linear segment. 

\subsection{Optical Spectrum}

\begin{figure}
	\includegraphics[width=\columnwidth]{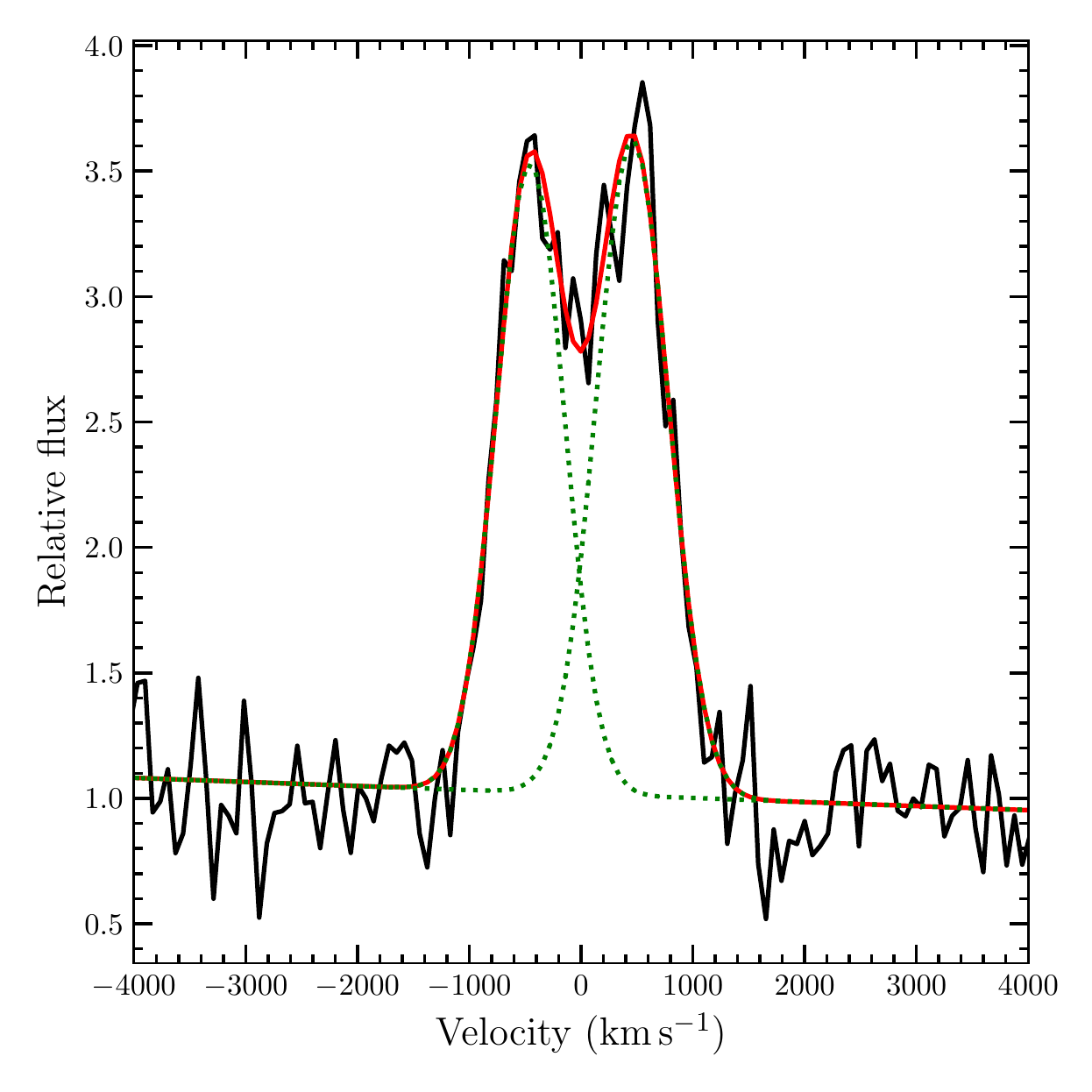}
	\caption{A Double Gaussian fit to the rotationally broadened H$_\alpha$ profile, obtained with the Gemini GMOS spectrograph on MJD~58324.3. The velocity range corresponds to a wavelength range of 6476--6651~\AA. The two Gaussian components have means of $+460\pm70\rm ~km\,s^{-1}$ and $-450\pm70\rm ~km\,s^{-1}$.   \label{fig:velocity}}
\end{figure}

The GMOS spectra revealed a double-peaked H$_\alpha$ emission line, as seen in Fig.~\ref{fig:velocity}. Other spectral features could not be recovered from the GMOS-N spectrum (see Fig.~\ref{fig:full_spec}). The double-peaked H$_\alpha$ emission line is produced by orbiting gas in the accretion disk seen at a significant inclination to our line of sight. To recover the observed disk velocity, we fit the data using $\chi^2$ minimisation with a model consisting of a double Gaussian added to a linearly varying continuum. As the 900s integration time is only a fraction of the expected $\sim80$~minute orbital period, we can not separate the systemic velocity of the binary from the orbital motion of the WD.
As seen in Fig.~\ref{fig:velocity} the redshifted H$_\alpha$ component was found to peak at $+450\pm60\rm ~km\,s^{-1}$ and the blue at $-460\pm60\rm ~km\,s^{-1}$ relative to the centre of the emission. The full width at half maximum (FWHM) of each of the Gaussian components was found to be $740\pm 45~\rm km\,s^{-1}$.

From the spectrum, we can roughly constrain system properties, such as inclination and WD mass. As discussed in \citet{Warner1986}, the equivalent width (EW) of the $H_\beta$ emission line is inclination dependent. Expanding on this, \citet{casares2015} has calibrated the EW and FWHM of the $H_\alpha$ emission line in CVs and X-ray transients to permit the estimate of inclination and mass. To first order, \citet{casares2015} make the following approximation:
\begin{eqnarray}
EW \approx \frac{B}{\textrm{cos}~i}
\end{eqnarray}
where $EW$ is the equivalent width, $B=9\pm 8$~\AA, and $i$ is the inclination. The value of the $B$ parameter is calibrated from a set of well-observed CVs. From our spectrum, we find the $H_\alpha$ $EW = -88\pm 4$, which leads to an inclination $i=84\pm 5^\circ$. We conclude that the inclination is fairly high, although it must be on the low side of this range as the system is not eclipsing.




\section{Discussion} \label{sec:discussion}

\subsection{Orbital Period and Mass Ratio}

\begin{figure}
	\includegraphics[width=\columnwidth]{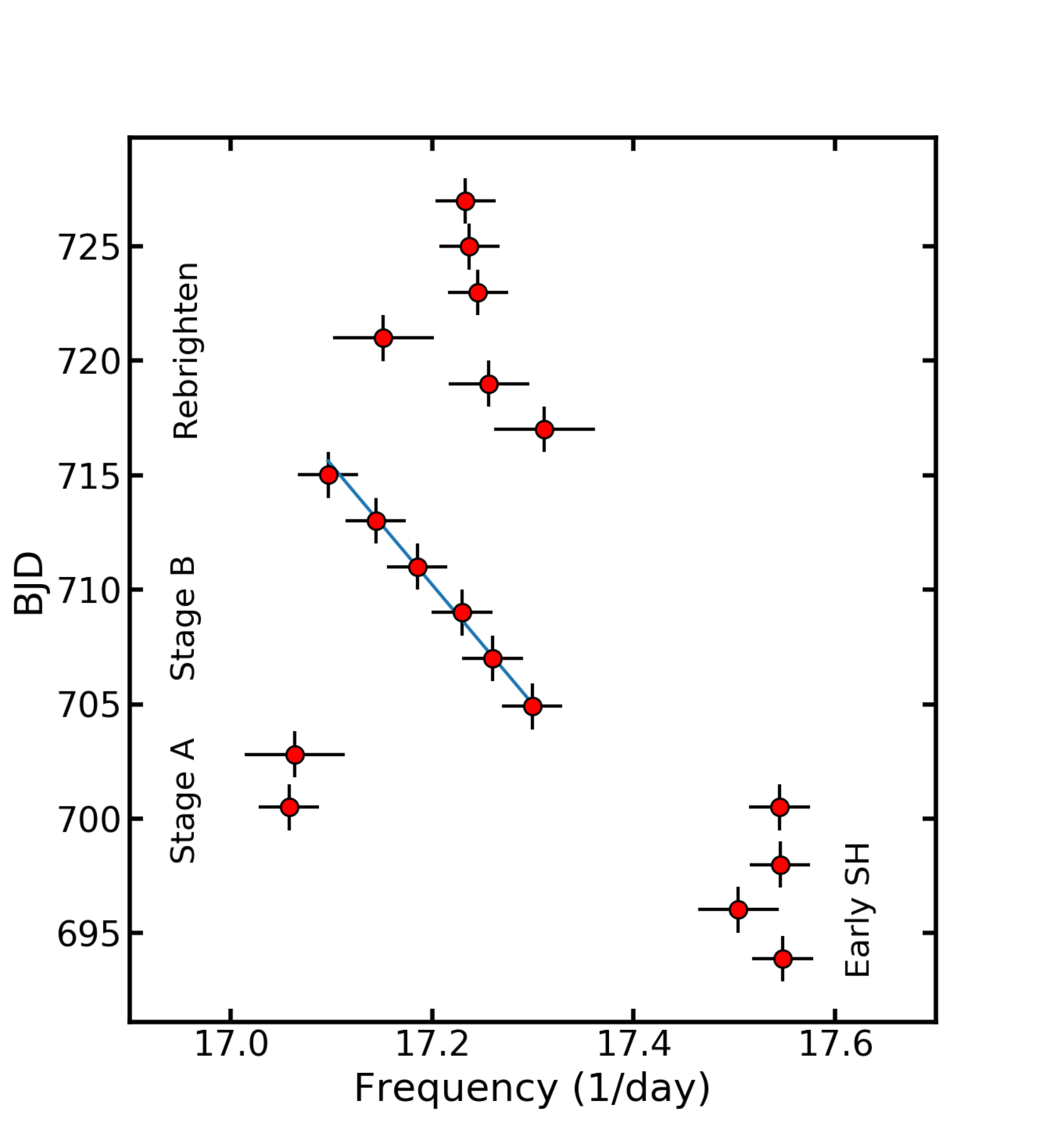}
	\caption{The frequency of the peak of power spectrum binned in two day time steps. The power spectrum centred on day 700.5 shows two strong peaks, so both frequencies are plotted. The first 9 days after peak brightness display ``early superhump'' oscillations that generally match the binary orbital period. Stage~A superhumps transition quickly to the Stage~B oscillations. The superhump period over Stage~B shifts to lower frequencies  The measured rate of period change is $\dot{P}=6.4\times 10^{-5}$. \label{fig:pdot}}
\end{figure}

Early superhumps are known to have a period within 0.1\%\ of the binary orbital period \citep{Kato2015}. The power spectrum starting from the peak of the light curve to day 702.4 (the onset of Stage~A superhumps) provides a good estimate of the orbital period of 82.13$\pm 0.03$ minutes (Fig.~\ref{fig:pdot}). As compiled by \citet{Kato2015}, the median orbital period of WZ~Sge stars is 81.94~minutes.

Ordinary superhumps are seen to develop at day 702 as the amplitude of oscillations rises from a few percent up to 8\% on day 703. The delay in the start of true superhumps is 
$9.8\pm 0.2$ days, which is typical for WZ~Sge stars with an 82~minute period.
Stage~A superhumps last only 20 to 30 cycles and are difficult to measure due to our 30~minute cadence. We apply a power spectrum analysis to the data between day 702.4 and 703.6 (20 cycles) and find a peak centred at $P_{SH}=84.40\pm 0.05$~minutes. Although, it is clear from the time-resolved power spectrum shown in Fig.~\ref{fig:power}, that the Stage~A superhumps evolve rapidly from lower frequencies toward the Stage~B frequencies. Fig.~\ref{fig:pdot} also shows that the early superhumps and the Stage~A oscillations are present simultaneously in the system just before the onset of Stage~B.  

\begin{figure}
	\includegraphics[width=\columnwidth]{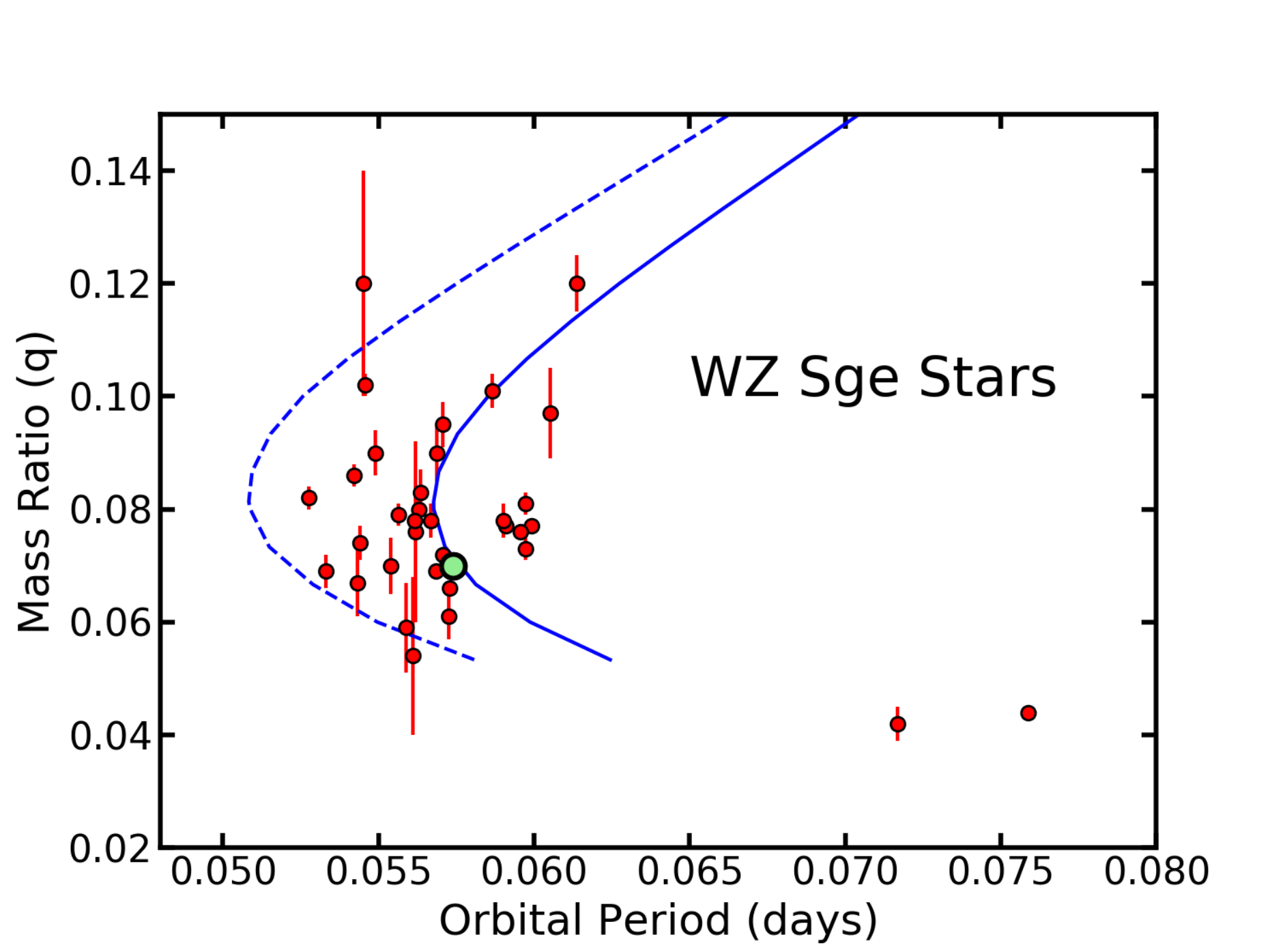}
	\caption{The estimated mass ratio, $q$, versus orbital period for the known WZ~Sge stars from \citet{Kato2015} (red points). \name\ is shown as a green point just after bounce. The dashed line is the standard evolutionary track for CVs from \citet{Knigge2011} while the solid line is their optimal binary track.  \label{fig:evolution}}
\end{figure}

The difference in period between the Stage~A superhumps and the orbital period is related to the binary mass ratio through the superhump excess parameter $\epsilon$, where $\epsilon = P_{SH}/P_{orb}-1$. We do not have an independent measurement of the orbital period, so we use the frequency of the early superhumps to estimate $\epsilon$. From $P_{SH}=84.40$~minutes and $P_{orb}=82.13$~minutes we find $\epsilon = 0.0276\pm 0.0005$, meaning the precession rate of the disk is $\epsilon^*=\epsilon/(1+\epsilon) = 0.0269$. \citet{kato2013b} determined a relation between the superhump excess and the components mass ratio and from their Table~1 the mass ratio for \name\ is $q=0.070\pm 0.005$. The orbital period and mass ratio places \name\ very close to the period bounce (Fig.~\ref{fig:evolution}).

The rate that the oscillation period changes with time during Stage~B superhumps is also a key diagnostic of the binary parameters. Fig.~\ref{fig:pdot} shows how the power peak trends to lower frequencies during Stage~B superhumps. We estimate the rate at which the superhump power spectrum peak shifts to lower frequencies to be $\dot{\nu} = -0.0191$ day$^{-2}$ corresponding to a period change rate of $\dot{P}=6.4\pm 0.5\times 10^{-5}$. For a collection of WZ~Sge stars with mass ratios between $0.06<q<0.08$ the measured period derivative ranges between $2\times 10^{-5} < \dot{P} <7\times 10^{-5}$ with a mean of 4$\times 10^{-5}$ \citep{Kato2015}. 

\citet{casares2016} has identified a relation between the quiescent H$_\alpha$ emission line properties and the binary mass ratio for systems with $q<0.25$.  These extremely low mass ratios are found in black hole X-ray binaries as well as CVs near the period bounce. \citet{casares2016} showed that the ratio between the separation in the double peaked $H_\alpha$ emission ($DP$) and the full width at half maximum (FWHM) of the two components is independent of the inclination of the disk, but sensitive to the mass ratio. The model predicts the relation,
\begin{eqnarray}
f(q) &=& \frac{0.49\left(1+q\right)^{-1}}{0.6+q^{2/3}ln\left(1+q^{-1/3} \right)}\\
\frac{DP}{FWHM}  &=& 3^{1/3}\left(1+q\right)^{2/3}\beta\sqrt{\alpha f\left(q\right)} \label{eq:dpfwhm}
\end{eqnarray}
where $\alpha$ and $\beta$ are free parameters. \citet{casares2016} estimated that $\alpha\approx 0.42$ and $\beta\approx 0.83$ by fitting Equation~\ref{eq:dpfwhm} to a set of known CVs. 

\begin{figure}
	\includegraphics[width=\columnwidth]{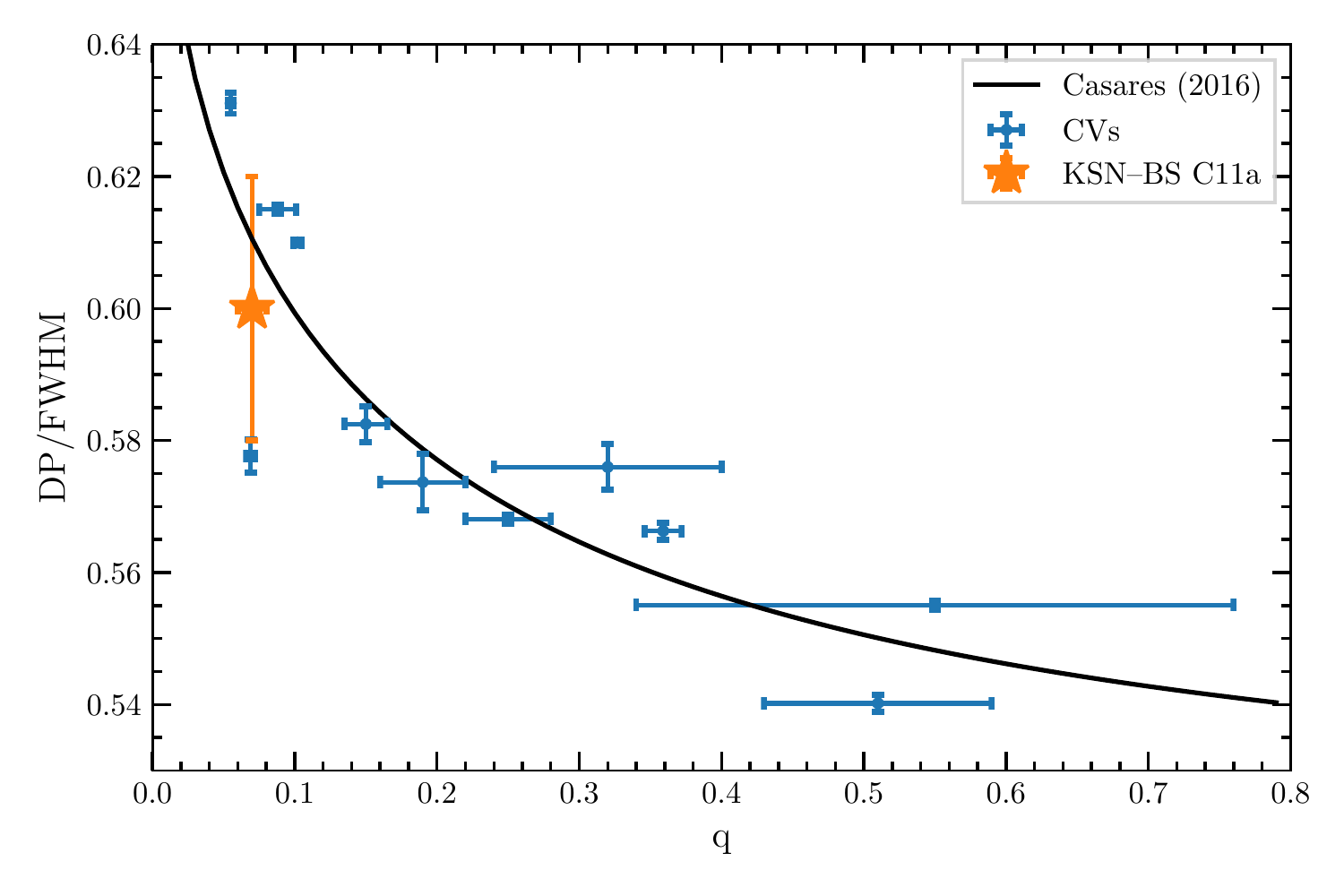}
	\caption{The H$_\alpha$ line ratio versus binary mass ratio, $q$, based on Fig.~2 from \citet{casares2016} for CVs. We add \name\ to the figure, which is consistent with the model parameters presented in \citet{casares2016} of $\alpha = 0.42$ and $\beta = 0.83$.
		\label{fig:dpfwhm}}
\end{figure}

From our Gemini spectrum of \name\, we measure $DP/FWHM = 0.60\pm 0.02$, consistent with a mass ratio of $q < 0.2$. Using our estimated mass ratio of $q = 0.070\pm0.005$, we can plot \name\ on the \citet{casares2016} relation for CVs as shown in Fig.~\ref{fig:dpfwhm}. The small mass ratio we have estimated for \name\ is consistent with this analysis, although, the scatter in the CV relation is large.

\subsection{Rebrightening Phase}

WZ~Sge stars show a variety of behaviours toward the end of a superoutburst and are generally referred to as ``rebrightening''. These behaviours have been divided into classes by \citet{Imada2006} and reviewed by \citet{Kato2015}. The rebrightening phase for \name\ begins with a slight rise followed by a shallow dip and then keeps a nearly constant brightness over five days before a final fade. The lack of large oscillations rules out the rebrightening classes of Type~B and C. \name\ also does not fit the Type~D rebrightening class characterised by GW~Lib. The behaviour of \name\ at the end of its outburst is probably closest to the Type~A class, like AL~Com, but with only a shallow dip before the plateau phase. The slight rise around day 717 suggests the rebrightening may have started before fading from the superoutburst began. This odd situation suggests that the fading begins in a different part of the disk than the rebrightening. Superhumps are seen to reform and intensify during the rebrightening phase.

\subsection{Early Rise}

\begin{figure*}
	\includegraphics[width=0.49\textwidth]{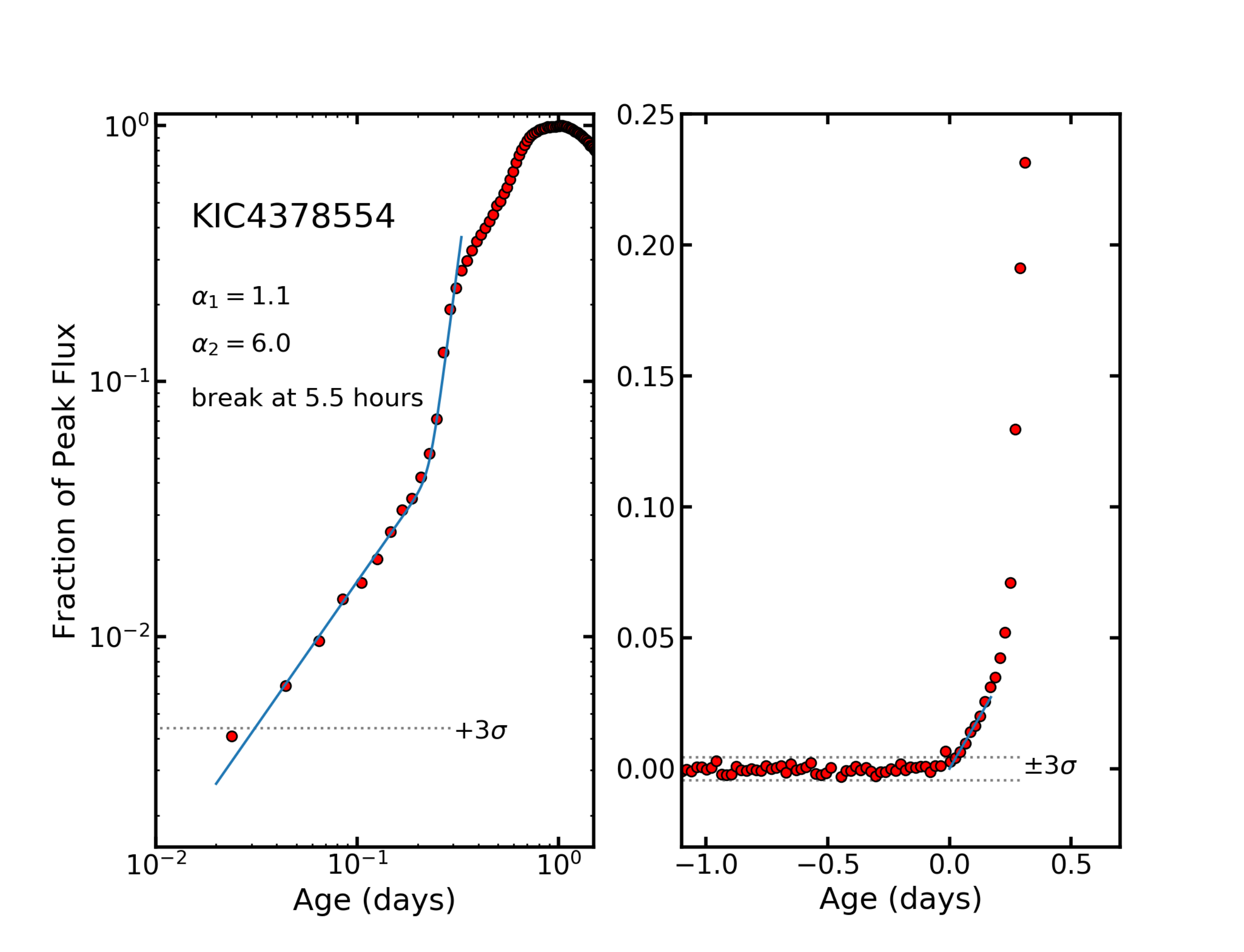}
	\includegraphics[width=0.49\textwidth]{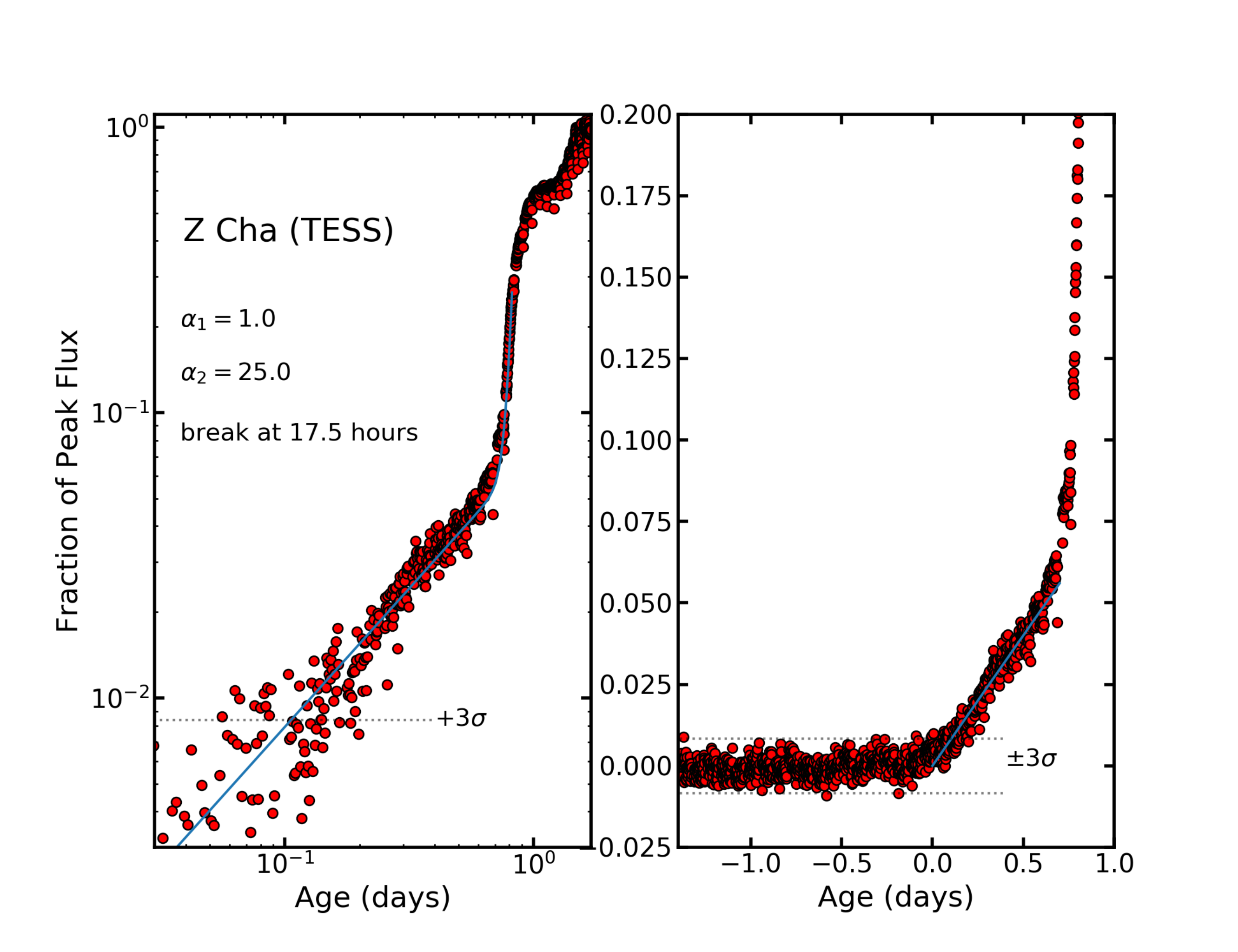}
	\caption{Early rises of superoutbursts from \citet{Barclay2012} (left) and Z~Cha (right). Since this feature is present in different dwarf novae types and different instruments, we expect this to be an indication of new physics in dwarf novae superoutbursts. \label{fig:others}}
\end{figure*}

The continuous monitoring and fast cadence of the \kepler\ and \textit{TESS} missions allows us to study the earliest phases of the superoutburst in SU~UMa stars. We see in Fig.~\ref{fig:rise} that \name\ shows a nearly linear precursor before the very fast rise to maximum light begins.
This prompted us to look at other short period systems for evidence of precursor behaviour. We found slow linear precursors are seen in the serendipitiously discovered system KIC~4378554 \citep{Barclay2012} as seen in Fig.~\ref{fig:others}. We find a similar broken rise in and Z~Cha from \textit{TESS} \citep{Court2019}. The detection of slow, linear precursors in these three objects suggests that such rises may be a common feature of SU~UMa outbursts.

The linear precursor in \name\ transitions to the fast rise at only 0.5\%\ of the peak flux of its outburst and lasts for 10~hours, or only 7 binary orbits. The linear phase of KIC~4378554 reaches 4\%\ of the peak flux but lasts for only 5.5~hours. The linear precursor in Z~Cha is most impressive as it reaches to 5\%\ of peak and lasts for 17.5~hours.  

%
%

If we assume the \textit{Kepler} bandpass captured most of the emission during the rise (bolometric flux), then the linear rise might suggest the temperature increased with time as $t^{1/4}$. A temperature increase of less than a factor of two could explain the factor of ten increase in flux during the early phase, pushing the disk to the critical ionization temperature for the neutral hydrogen. The temperature of quiescent disks vary with radius, but typical estimates range between 3000~K to 5600~K \citep{rutkowski2016}. So a rise of a factor of two in temperature is reasonable at the onset of an outburst.

\subsection{Distance}

The typical luminosity of WZ~Sge stars in quiescence is not well determined because these objects are so intrinsically faint and only a handful have apparent magnitude brighter than 19 in quiescence. The physical origin of the ordinary superhumps suggests a fairly uniform disk luminosity for WZ~Sge systems when these oscillations begin \citep{Kato2015}. \citet{Patterson2011} estimates the absolute magnitude for SU~UMa stars near the time of superhump onset to be $M_V =5.5\pm 0.2$~mag, so we can use this as an indicator of the distance to \name.

The apparent \textit{Kepler} magnitude at superhump appearance is $K_p =15.4~mag$. The quiescent mag was 21.5, so the amplitude at the time at the onset of ordinary superhumps is 6.1~mag, typical of WZ~Sge stars \citep{Kato2015}. We do not know a precise inclination for \name, but it is not an eclipsing system so we will assume no significant inclination correction is required. The distance modulus to \name\ is $m-M= 9.9$ mag, or $\sim 900$~pc. The Galactic coordinates \name\ are $l=357^\circ$ and $b=12^\circ$, extremely close to the Galactic center. The total dust extinction in the direction of \name\ is a substantial $A_V=1.03$ \citep{Schlafly2011}, arguing for a distance as close as 600~pc. Given the relatively blue colour of the star in quiescence, we suspect that much of the extinction lies beyond \name, and the true distance is closer to 900~pc.

\section{Conclusion}

Searching through the \kepler\ pixels files as part of the K2:Background Survey, we have identified a bright transient. The continuous monitoring and rapid cadence of \kepler\ showed that the transient increased in brightness by 8~mag in about one day and faded slowly over a month. Oscillations with varying frequencies strongly suggest that the transient is a WZ~Sge type dwarf nova with a binary orbital period of 82~minutes. This type of close binary lose energy by emitting gravitational radiation. The stars get closer together driving continued mass transfer mass to the WD. But mass transfer forces the secondary out of thermal equilibrium and around a period of 80~min the period ``bounces'' so that evolution continues with an increasing period \citep{Knigge2011}.

We identify the progenitor system in quiescence using spectroscopy and deep DECam imaging. The spectra reveal a double-peaked H$_\alpha$ emission line coming from the quiescent accretion disk. Its equivalent width suggests a high inclination system, although, eclipses are not seen in the outbursting light curve. The line properties are consistent with a very low mass ratio as expected for a WZ~Sge system.  

We also find that SU-UMa DN superoutbursts exhibit a broken power law rise. This phenomena has been observed in \name, KIC4378554, and Z~Cha. The mechanism behind this broken rise is unknown, however, we expect it indicates new physics behind DN superoutbursts.

\section*{Acknowledgements}
This research weas supported by an Australian Government Research Training Program (RTP) Scholarship. This paper includes data collected by the K2 mission. Funding for the K2 mission is provided by the NASA Science Mission directorate. Based on observations obtained at the Gemini Observatory, as part of the GN-2018A-LP-14 program, which is operated by the Association of Universities for Research in Astronomy, Inc., under a cooperative agreement with the NSF on behalf of the Gemini partnership: the National Science Foundation (United States), National Research Council (Canada), CONICYT (Chile), Ministerio de Ciencia, Tecnolog\'{i}a e Innovaci\'{o}n Productiva (Argentina), Minist\`{e}rio da Ci\^{e}ncia, Tecnologia e Inova\c{c}\~{a}o (Brazil), and Korea Astronomy and Space Science Institute (Republic of Korea).


\begin{table}
\caption{Proposals to observe EPIC 203830112, all are independent of this paper. \label{tab:proposals}}
\begin{tabular}{lll}
\hline
Proposal ID & Campaign & PI\\
\hline
 GO2054 & C02 & Sanchis-Ojeda et al. \\
 GO2104 & C02 & Petigura et al. \\
 GO11071 & C11 & Charbonneau et al. \\
 GO11122 & C11 & Howard et al. \\
 GO11902 & C11 & GO Office, K2 \\
 \hline
 \end{tabular}
\end{table}

\begin{table}
\caption{\name\ Characteristics \label{table1}}
\begin{tabular}{llll}
\hline
Parameter & Value & Uncertainty & Units \\
\hline
Outburst type & A & &  \\
RA                        & 16:53:50.67   &  0.24''     & J2000 \\
DEC                       & -24:46:26.50  &  0.24''     & J2000 \\
Time of outburst & 2457690.66 & 0.01 & BJD \\ 
Break in rise & 10.0 & 0.2 & hours \\ 
Full Outburst Amplitude   & $\sim 8.1$    & 0.2   & mag \\
Orbital Period$^a$        & 0.05704       & 0.1\% & days   \\
Stage A Period            & 0.0586        & 0.2\% & days   \\
Stage B, $\dot{\nu}$       & $-0.0191$     & 2.6\% & day$^{-2}$  \\
SH Excess, $\epsilon$     & 0.0276        & 1.8\% &    \\
Mass Ratio, $q$           & 0.070         & 0.01  & \\ 
Disk velocity, $v$       & 460           & 70    & $\rm km\,s^{-1}$ \\
Orbital inclination, $i$ &       84        & 5       & deg \\
Distance &       $\sim 900$        &        & pc \\
\hline
\multicolumn{3}{l}{$^a$ From Early Superhumps}
\end{tabular}

\end{table}

\appendix

\section{GMOS-N spectrum} \label{sec:full_spec}
\begin{figure*}
	\includegraphics[width=\textwidth]{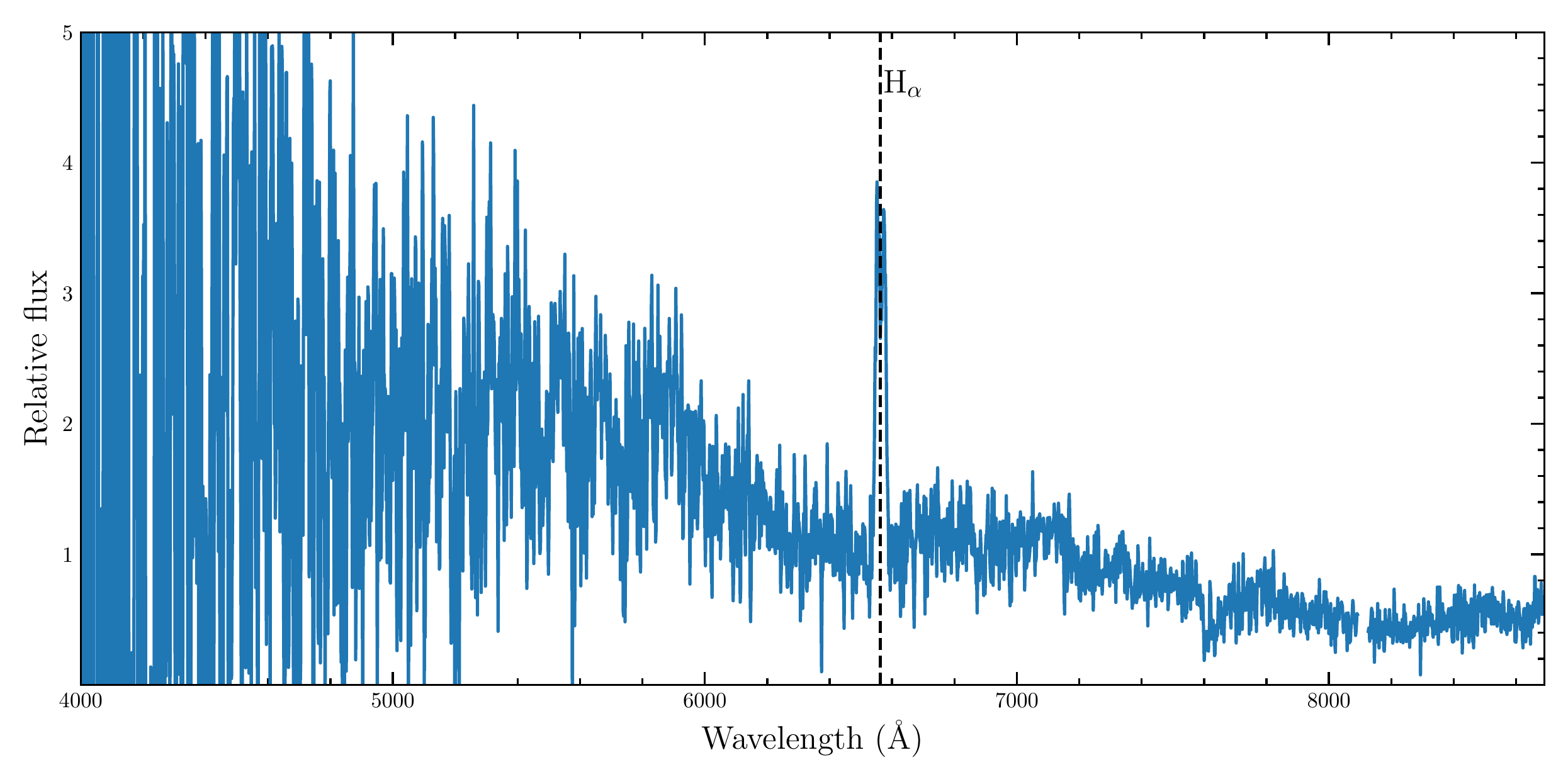}
	\caption{The entire wavelength range of the GMOS-N 900~s spectrum taken on MJD58324.3. The H$_\alpha$ emission is indicated by the dashed vertical line, and the A band telluric absorption line is a seen from 7600--7650 \AA.  \label{fig:full_spec}}
\end{figure*}

The full range of the GMOS-N spectrum is shown in Fig.~\ref{fig:full_spec}. As the spectrum was taken while \name\ was in quiescence, the source is faint, with only the H$_\alpha$ emission clearly resolved. The A band telluric absorption line is seen from 7600--7650 \AA.

\bibliographystyle{mnras}
\bibliography{AllPhD.bib}

\begin{thebibliography}{}
\makeatletter
\relax
\def\mn@urlcharsother{\let\do\@makeother \do\$\do\&\do\#\do\^\do\_\do\%\do\~}
\def\mn@doi{\begingroup\mn@urlcharsother \@ifnextchar [ {\mn@doi@}
  {\mn@doi@[]}}
\def\mn@doi@[#1]#2{\def\@tempa{#1}\ifx\@tempa\@empty \href
  {http://dx.doi.org/#2} {doi:#2}\else \href {http://dx.doi.org/#2} {#1}\fi
  \endgroup}
\def\mn@eprint#1#2{\mn@eprint@#1:#2::\@nil}
\def\mn@eprint@arXiv#1{\href {http://arxiv.org/abs/#1} {{\tt arXiv:#1}}}
\def\mn@eprint@dblp#1{\href {http://dblp.uni-trier.de/rec/bibtex/#1.xml}
  {dblp:#1}}
\def\mn@eprint@#1:#2:#3:#4\@nil{\def\@tempa {#1}\def\@tempb {#2}\def\@tempc
  {#3}\ifx \@tempc \@empty \let \@tempc \@tempb \let \@tempb \@tempa \fi \ifx
  \@tempb \@empty \def\@tempb {arXiv}\fi \@ifundefined
  {mn@eprint@\@tempb}{\@tempb:\@tempc}{\expandafter \expandafter \csname
  mn@eprint@\@tempb\endcsname \expandafter{\@tempc}}}

\bibitem[\protect\citeauthoryear{{Barclay}, {Still}, {Jenkins}, {Howell}  \&
  {Roettenbacher}}{{Barclay} et~al.}{2012}]{Barclay2012}
{Barclay} T.,  {Still} M.,  {Jenkins} J.~M.,  {Howell} S.~B.,   {Roettenbacher}
  R.~M.,  2012, \mn@doi [\mnras] {10.1111/j.1365-2966.2012.20700.x}, \href
  {https://ui.adsabs.harvard.edu/abs/2012MNRAS.422.1219B} {422, 1219}

\bibitem[\protect\citeauthoryear{{Borucki} et~al.,}{{Borucki}
  et~al.}{2010}]{Borucki2010}
{Borucki} W.~J.,  et~al., 2010, \mn@doi [Science] {10.1126/science.1185402},
  \href {https://ui.adsabs.harvard.edu/abs/2010Sci...327..977B} {327, 977}

\bibitem[\protect\citeauthoryear{{Brown} et~al.,}{{Brown}
  et~al.}{2015}]{Brown2015b}
{Brown} A.,  et~al., 2015, \mn@doi [\aj] {10.1088/0004-6256/149/2/67}, \href
  {https://ui.adsabs.harvard.edu/abs/2015AJ....149...67B} {149, 67}

\bibitem[\protect\citeauthoryear{{Casares}}{{Casares}}{2015}]{casares2015}
{Casares} J.,  2015, \mn@doi [\apj] {10.1088/0004-637X/808/1/80}, \href
  {https://ui.adsabs.harvard.edu/abs/2015ApJ...808...80C} {808, 80}

\bibitem[\protect\citeauthoryear{{Casares}}{{Casares}}{2016}]{casares2016}
{Casares} J.,  2016, \mn@doi [\apj] {10.3847/0004-637X/822/2/99}, \href
  {https://ui.adsabs.harvard.edu/abs/2016ApJ...822...99C} {822, 99}

\bibitem[\protect\citeauthoryear{{Chambers} et~al.,}{{Chambers}
  et~al.}{2016}]{Chambers2016}
{Chambers} K.~C.,  et~al., 2016, arXiv e-prints, \href
  {https://ui.adsabs.harvard.edu/abs/2016arXiv161205560C} {p. arXiv:1612.05560}

\bibitem[\protect\citeauthoryear{{Court} et~al.,}{{Court}
  et~al.}{2019}]{Court2019}
{Court} J.~M.~C.,  et~al., 2019, \mn@doi [\mnras] {10.1093/mnras/stz2015},
  \href {https://ui.adsabs.harvard.edu/abs/2019MNRAS.488.4149C} {488, 4149}

\bibitem[\protect\citeauthoryear{{Dimitriadis} et~al.,}{{Dimitriadis}
  et~al.}{2019}]{Dimitriadis2018}
{Dimitriadis} G.,  et~al., 2019, \mn@doi [\apjl] {10.3847/2041-8213/aaedb0},
  \href {https://ui.adsabs.harvard.edu/abs/2019ApJ...870L...1D} {870, L1}

\bibitem[\protect\citeauthoryear{{Garnavich}, {Tucker}, {Rest}, {Shaya},
  {Olling}, {Kasen}  \& {Villar}}{{Garnavich} et~al.}{2016}]{Garnavich2016}
{Garnavich} P.~M.,  {Tucker} B.~E.,  {Rest} A.,  {Shaya} E.~J.,  {Olling}
  R.~P.,  {Kasen} D.,   {Villar} A.,  2016, \mn@doi [\apj]
  {10.3847/0004-637X/820/1/23}, \href
  {https://ui.adsabs.harvard.edu/abs/2016ApJ...820...23G} {820, 23}

\bibitem[\protect\citeauthoryear{{H{\={o}}shi}}{{H{\={o}}shi}}{1979}]{Hoshi1979}
{H{\={o}}shi} R.,  1979, \mn@doi [Progress of Theoretical Physics]
  {10.1143/PTP.61.1307}, \href
  {https://ui.adsabs.harvard.edu/abs/1979PThPh..61.1307H} {61, 1307}

\bibitem[\protect\citeauthoryear{{Howell} et~al.,}{{Howell}
  et~al.}{2014}]{Howell2014}
{Howell} S.~B.,  et~al., 2014, \mn@doi [\pasp] {10.1086/676406}, \href
  {https://ui.adsabs.harvard.edu/abs/2014PASP..126..398H} {126, 398}

\bibitem[\protect\citeauthoryear{{Imada}, {Kato}, {Monard}, {Retter}, {Liu}  \&
  {Nogami}}{{Imada} et~al.}{2006}]{Imada2006}
{Imada} A.,  {Kato} T.,  {Monard} L.~A.~G.,  {Retter} A.,  {Liu} A.,   {Nogami}
  D.,  2006, \mn@doi [\pasj] {10.1093/pasj/58.2.383}, \href
  {https://ui.adsabs.harvard.edu/abs/2006PASJ...58..383I} {58, 383}

\bibitem[\protect\citeauthoryear{{Ishioka} et~al.,}{{Ishioka}
  et~al.}{2002}]{ishioka2002}
{Ishioka} R.,  et~al., 2002, \mn@doi [\aap] {10.1051/0004-6361:20011644}, \href
  {https://ui.adsabs.harvard.edu/abs/2002A&A...381L..41I} {381, L41}

\bibitem[\protect\citeauthoryear{{Kato}}{{Kato}}{2002}]{kato2002}
{Kato} T.,  2002, \mn@doi [\pasj] {10.1093/pasj/54.2.L11}, \href
  {https://ui.adsabs.harvard.edu/abs/2002PASJ...54L..11K} {54, L11}

\bibitem[\protect\citeauthoryear{{Kato}}{{Kato}}{2015}]{Kato2015}
{Kato} T.,  2015, \mn@doi [\pasj] {10.1093/pasj/psv077}, \href
  {https://ui.adsabs.harvard.edu/abs/2015PASJ...67..108K} {67, 108}

\bibitem[\protect\citeauthoryear{{Kato} \& {Osaki}}{{Kato} \&
  {Osaki}}{2013a}]{Kato2013a}
{Kato} T.,  {Osaki} Y.,  2013a, \mn@doi [\pasj] {10.1093/pasj/65.5.97}, \href
  {https://ui.adsabs.harvard.edu/abs/2013PASJ...65...97K} {65, 97}

\bibitem[\protect\citeauthoryear{{Kato} \& {Osaki}}{{Kato} \&
  {Osaki}}{2013b}]{kato2013b}
{Kato} T.,  {Osaki} Y.,  2013b, \mn@doi [\pasj] {10.1093/pasj/65.6.115}, \href
  {https://ui.adsabs.harvard.edu/abs/2013PASJ...65..115K} {65, 115}

\bibitem[\protect\citeauthoryear{{Kato} \& {Osaki}}{{Kato} \&
  {Osaki}}{2013c}]{Kato2013}
{Kato} T.,  {Osaki} Y.,  2013c, \mn@doi [\pasj] {10.1093/pasj/65.6.L13}, \href
  {https://ui.adsabs.harvard.edu/abs/2013PASJ...65L..13K} {65, L13}

\bibitem[\protect\citeauthoryear{{Kato} et~al.,}{{Kato}
  et~al.}{2014}]{kato2014}
{Kato} T.,  et~al., 2014, \mn@doi [\pasj] {10.1093/pasj/psu072}, \href
  {https://ui.adsabs.harvard.edu/abs/2014PASJ...66...90K} {66, 90}

\bibitem[\protect\citeauthoryear{{Knigge}, {Baraffe}  \& {Patterson}}{{Knigge}
  et~al.}{2011}]{Knigge2011}
{Knigge} C.,  {Baraffe} I.,   {Patterson} J.,  2011, \mn@doi [\apjs]
  {10.1088/0067-0049/194/2/28}, \href
  {https://ui.adsabs.harvard.edu/abs/2011ApJS..194...28K} {194, 28}

\bibitem[\protect\citeauthoryear{{Koch} et~al.,}{{Koch}
  et~al.}{2010}]{Koch2010}
{Koch} D.~G.,  et~al., 2010, \mn@doi [\apjl] {10.1088/2041-8205/713/2/L79},
  \href {https://ui.adsabs.harvard.edu/abs/2010ApJ...713L..79K} {713, L79}

\bibitem[\protect\citeauthoryear{{Kolb} \& {Baraffe}}{{Kolb} \&
  {Baraffe}}{1999}]{kolb1999}
{Kolb} U.,  {Baraffe} I.,  1999, \mn@doi [\mnras]
  {10.1046/j.1365-8711.1999.02926.x}, \href
  {https://ui.adsabs.harvard.edu/abs/1999MNRAS.309.1034K} {309, 1034}

\bibitem[\protect\citeauthoryear{{Lightkurve Collaboration}
  et~al.,}{{Lightkurve Collaboration} et~al.}{2018}]{lightkurve}
{Lightkurve Collaboration} et~al., 2018, {Lightkurve: Kepler and TESS time
  series analysis in Python}, Astrophysics Source Code Library (\mn@eprint
  {ascl} {1812.013})

\bibitem[\protect\citeauthoryear{{Lubow}}{{Lubow}}{1991a}]{Lubow1991}
{Lubow} S.~H.,  1991a, \mn@doi [\apj] {10.1086/170647}, \href
  {https://ui.adsabs.harvard.edu/abs/1991ApJ...381..259L} {381, 259}

\bibitem[\protect\citeauthoryear{{Lubow}}{{Lubow}}{1991b}]{Lubow1991a}
{Lubow} S.~H.,  1991b, \mn@doi [\apj] {10.1086/170648}, \href
  {https://ui.adsabs.harvard.edu/abs/1991ApJ...381..268L} {381, 268}

\bibitem[\protect\citeauthoryear{{Osaki}}{{Osaki}}{1974}]{Osaki1974}
{Osaki} Y.,  1974, \pasj, \href
  {https://ui.adsabs.harvard.edu/abs/1974PASJ...26..429O} {26, 429}

\bibitem[\protect\citeauthoryear{Osaki}{Osaki}{1989}]{Osaki1989}
Osaki Y.,  1989, Publications of the Astronomical Society of Japan, 41, 1005

\bibitem[\protect\citeauthoryear{{Osaki} \& {Meyer}}{{Osaki} \&
  {Meyer}}{2002}]{Osaki2002}
{Osaki} Y.,  {Meyer} F.,  2002, \mn@doi [\aap] {10.1051/0004-6361:20011744},
  \href {https://ui.adsabs.harvard.edu/abs/2002A&A...383..574O} {383, 574}

\bibitem[\protect\citeauthoryear{{Patterson}}{{Patterson}}{2011}]{Patterson2011}
{Patterson} J.,  2011, \mn@doi [\mnras] {10.1111/j.1365-2966.2010.17881.x},
  \href {https://ui.adsabs.harvard.edu/abs/2011MNRAS.411.2695P} {411, 2695}

\bibitem[\protect\citeauthoryear{{Rest} et~al.,}{{Rest}
  et~al.}{2018}]{Rest2018}
{Rest} A.,  et~al., 2018, \mn@doi [Nature Astronomy]
  {10.1038/s41550-018-0423-2}, \href
  {https://ui.adsabs.harvard.edu/abs/2018NatAs...2..307R} {2, 307}

\bibitem[\protect\citeauthoryear{Ridden-Harper, Tucker, Gully-Santiago,
  Barentsen, Rest, Garnavich  \& Shaya}{Ridden-Harper
  et~al.}{2020}]{Ridden-Harper2020}
Ridden-Harper R.,  Tucker B.~E.,  Gully-Santiago M.,  Barentsen G.,  Rest A.,
  Garnavich P.,   Shaya E.,  2020, \mn@doi [MNRAS] {10.1093/mnras/staa2247},
  12, 1

\bibitem[\protect\citeauthoryear{{Rutkowski}, {Waniak}, {Preston}  \&
  {Pych}}{{Rutkowski} et~al.}{2016}]{rutkowski2016}
{Rutkowski} A.,  {Waniak} W.,  {Preston} G.,   {Pych} W.,  2016, \mn@doi
  [\mnras] {10.1093/mnras/stw2139}, \href
  {https://ui.adsabs.harvard.edu/abs/2016MNRAS.463.3290R} {463, 3290}

\bibitem[\protect\citeauthoryear{{Schlafly} \& {Finkbeiner}}{{Schlafly} \&
  {Finkbeiner}}{2011}]{Schlafly2011}
{Schlafly} E.~F.,  {Finkbeiner} D.~P.,  2011, \mn@doi [\apj]
  {10.1088/0004-637X/737/2/103}, \href
  {https://ui.adsabs.harvard.edu/abs/2011ApJ...737..103S} {737, 103}

\bibitem[\protect\citeauthoryear{{Vican} et~al.,}{{Vican}
  et~al.}{2011}]{vican2011}
{Vican} L.,  et~al., 2011, \mn@doi [\pasp] {10.1086/662633}, \href
  {https://ui.adsabs.harvard.edu/abs/2011PASP..123.1156V} {123, 1156}

\bibitem[\protect\citeauthoryear{{Warner}}{{Warner}}{1986}]{Warner1986}
{Warner} B.,  1986, \mn@doi [\mnras] {10.1093/mnras/222.1.11}, \href
  {https://ui.adsabs.harvard.edu/abs/1986MNRAS.222...11W} {222, 11}

\bibitem[\protect\citeauthoryear{{Wenger} et~al.,}{{Wenger}
  et~al.}{2000}]{Wenger2000}
{Wenger} M.,  et~al., 2000, \mn@doi [\aaps] {10.1051/aas:2000332}, \href
  {https://ui.adsabs.harvard.edu/abs/2000A&AS..143....9W} {143, 9}

\bibitem[\protect\citeauthoryear{{Whitehurst}}{{Whitehurst}}{1988}]{Whitehurst1988}
{Whitehurst} R.,  1988, \mn@doi [\mnras] {10.1093/mnras/232.1.35}, \href
  {https://ui.adsabs.harvard.edu/abs/1988MNRAS.232...35W} {232, 35}

\makeatother
\end{thebibliography}

\bsp	
\label{lastpage}
\end{document}